\documentclass[11pt]{iopart}
\pdfoutput=1
\expandafter\let\csname equation*\endcsname\relax
\expandafter\let\csname endequation*\endcsname\relax
\usepackage[numbers,sort&compress]{natbib} 
\usepackage[pdftex]{graphicx,color}
\usepackage{amsmath,amsfonts,amssymb,wasysym,graphicx,capt-of,ifthen,calc}
\usepackage{enumerate,float,url,color}
\usepackage{amsmath}
\usepackage{graphics}  
\usepackage[latin1]{inputenc}
\usepackage{color}
\usepackage{rotating}
\usepackage{mathtools}
\begin{document}
\title{Transport properties of diffusive particles conditioned to survive in trapping environments}
\author{Gaia Pozzoli}
\address{Dipartimento di Scienza e Alta Tecnologia and Center for Nonlinear and Complex Systems, Universit\`a degli Studi dell'Insubria, Via Valleggio 11, 22100 Como Italy}
\address{I.N.F.N. Sezione di Milano, Via Celoria 16, 20133 Milano, Italy}
\author{Benjamin De Bruyne}
\address{LPTMS, CNRS, Univ.\ Paris-Sud, Universit\'e Paris-Saclay, 91405 Orsay, France}

\begin{abstract}
 We consider a one-dimensional Brownian motion with diffusion coefficient $D$ in the presence of $n$ partially absorbing traps with intensity $\beta$, separated by a distance $L$ and evenly spaced around the initial position of the particle. We study the transport properties of the process conditioned to survive up to time $t$. We find that the surviving particle first diffuses normally, before it encounters the traps, then undergoes a period of transient anomalous diffusion, after which it reaches a final diffusive regime. The asymptotic regime is governed by an effective diffusion coefficient $D_\text{eff}$, which is induced by the trapping environment and is typically different from the original one. We show that when the number of traps is \emph{finite}, the environment enhances diffusion and induces an effective diffusion coefficient that is systematically equal to $D_\text{eff}=2D$, independently of the number of the traps, the trapping intensity $\beta$ and the distance $L$. On the contrary, when the number of traps is \emph{infinite}, we find that the environment inhibits diffusion with an effective diffusion coefficient that depends on the traps intensity $\beta$ and the distance $L$ through a non-trivial scaling function $D_\text{eff}=D \mathcal{F}(\beta L/D)$, for which we obtain a closed-form. Moreover, we provide a rejection-free algorithm to generate surviving trajectories by deriving an effective Langevin equation with an effective repulsive potential induced by the traps. Finally, we extend our results to other trapping environments.
\end{abstract}

\section{Introduction}

\subsection{Introduction}
Death processes occur in many physical, biological, environmental and chemical reactions. Their transport and statistical properties are often modeled by diffusive particles in an environment composed of multiple traps. The survival probability, which is the probability that a particle did not get absorbed by one of the traps, is one of the most simple, yet difficult, observables to study and has attracted a sustained interest in the physics community  \cite{Lifshitz63,Lifshitz65,Lifshitz88,Balagurov74,Rosenstock70,Donsker75,Donsker79,Schuss1,Schuss2,Schuss3,Schuss4,Pinsky3}. This observable has been widely investigated in various trapping environments \cite{Bramson88,Bray02a,Bray02b,Bray02c,Majumdar03,Bray03,Yuste08,Krapivsky2014,Ledoussal2009,Texier2009,Grabsch2014,Condamin05} and has shown to have numerous applications in nature, such as in target searching strategies \cite{Oshanin02} and diffusion-limited reactions \cite{Smol16,Chand43,Benson60,Rice85,mortal-walkers}. 

In many practical circumstances, it is necessary to go beyond the survival probability and to analyse the transport properties of the system. Starting from a statistical ensemble of particles in the presence of a trapping environment, it is natural to focus on the surviving particles, i.e.~the ones that did not get absorbed, in order to derive the transport properties of the system. Since the proportion of surviving particles decreases with time, it is necessary to renormalize the probability density function by the remaining mass to recover a probability measure at all times. This is particularly relevant for a variety of situations such as in acoustics \cite{Mortessagne1993,Kuttruff}, optics \cite{ref-optics,ref-optics2}, resonant open quantum systems \cite{Nonnemnmacher2008}, and ``vicious walkers'' \cite{Grela2021,Fisher84}. While transport properties of absorbing systems are often described stochastically, they can also be studied deterministically in the context of open dynamical systems \cite{rev-det}, such as in chaotic billiards in the presence of holes in phase space \cite{Demers2005}, expanding maps on non-invariant sets \cite{Kaufmann1997, Pianigiani1979}, partial absorption in chaotic systems \cite{Altmann2013}, etc. In this setting, the analysis of the normalized distribution leads to the notions of escape rate and convergence to a \emph{conditionally invariant measure}.

Interestingly, the presence of a trapping environment sometimes yields to a regime of anomalous diffusion where the mean-square displacement of a diffusive particle is no longer proportional to time. The anomalous diffusion can either be an asymptotic property of the system or a transient non-equilibrium state with a finite lifetime, in which case it is usually referred to as \emph{transient anomalous diffusion} \cite{Ising-transient-subdiffusion,Yamamoto2015}. In the latter case, there exists a characteristic time at which the system switches over into a long term permanent dynamics. In the deterministic context, the analogous notion of transient chaos appears in systems that exhibit an irregular motion for some time and a subsequent transition to an asymptotic regular behavior (see for instance \cite{transient-chaos,transient-chaos-book}).

\begin{figure}
  \begin{center}
    \includegraphics[width=0.5\textwidth]{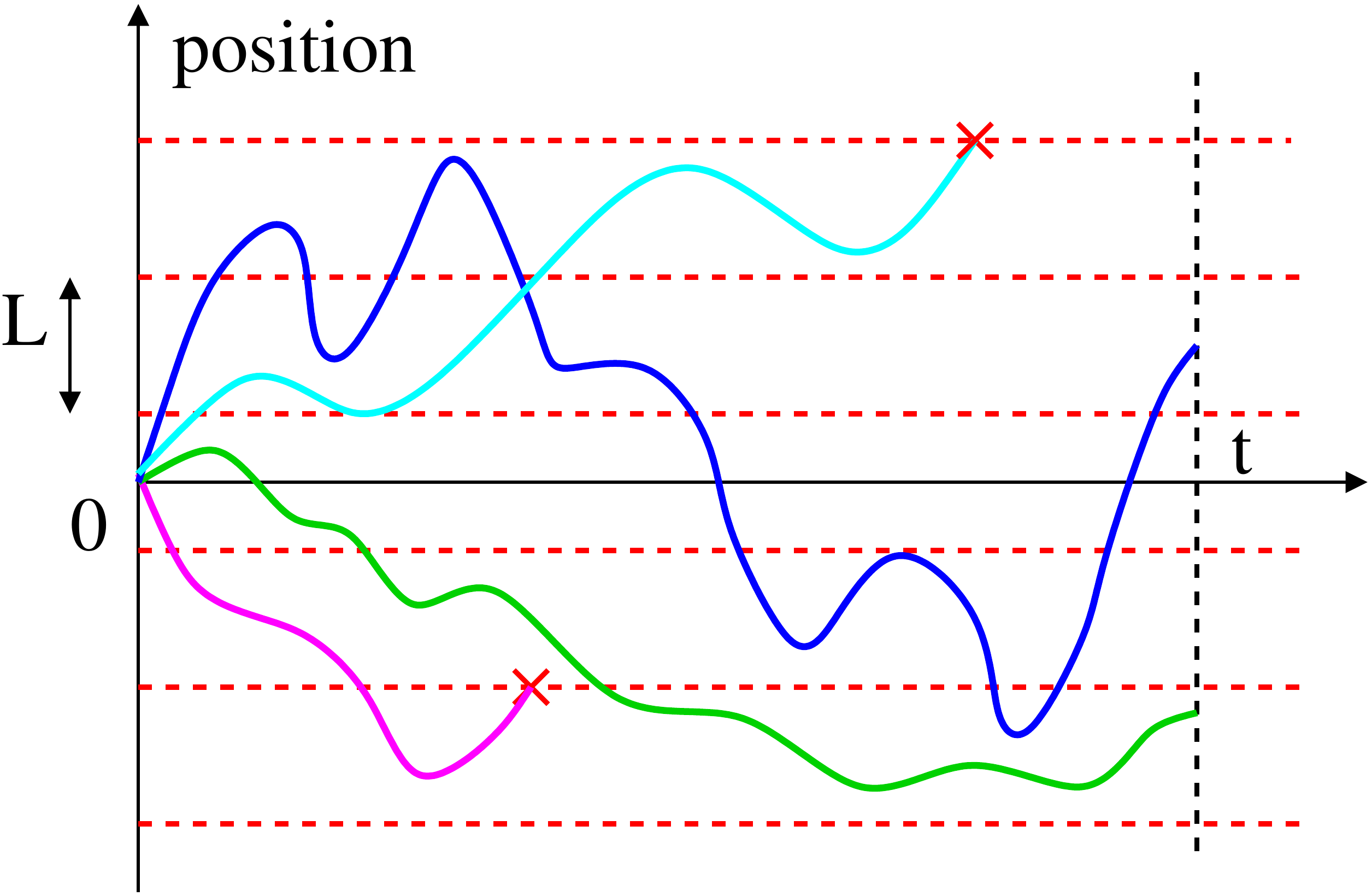}
    \caption{Schematic representation of trajectories of particles diffusing in the presence of $n=6$ partially absorbing traps (red dashed lines) with intensity $\beta$ separated by a distance $L$ and evenly spaced around the initial position of the particles. As the traps are partially absorbing, the particles can cross a trap several times before being absorbed. Within the four depicted trajectories, only two particles (green and blue) survived up to time $t$. The effective diffusion coefficient in (\ref{eq:DeffGen}) is computed by averaging only over the surviving trajectories at time $t$. We find that it is equal to $D_\text{eff}=2D$ when  the number of traps $n$ is finite as the surviving particles typically diffuse away towards $x\to \pm \infty$. When the number of traps is infinite $n\to \infty$, the surviving particles typically stay in-between the traps and the effective diffusion coefficient follows a non-trivial behavior $D_\text{eff}=\mathcal{F}(\beta L/D)$, where $\mathcal{F}$ is a scaling function given in (\ref{eq:Deffinf}).}
    \label{fig:model}
  \end{center}
\end{figure}

Recently, we have investigated a trapping environment composed of periodically distributed traps and highlighted the effect of a finite size of the traps on the survival probability \cite{GPBD21}. In the diffusive limit, we have found that the problem reduces to finding the survival probability of a Brownian motion with diffusion coefficient $D$ in the presence of partially absorbing traps with intensity $\beta$ and separated by a distance $L$. While the limit $\beta\to \infty$ corresponds to fully absorbing traps where the particle is absorbed upon its first encounter with a trap, the case of a finite $\beta$ corresponds to partially absorbing traps where the particle can cross a trap several times before being absorbed (see figure \ref{fig:model} with $n\to \infty$ number of point absorbers). 
In this work, we go beyond the survival probability and analyse the transport properties of this model, which have not yet been studied. In particular, we are interested in the effective diffusion coefficient $D_\text{eff}$ of the surviving particles, i.e.~
\begin{align}
  D_\text{eff} = \lim_{t\to \infty}\left[\frac{\langle x^2(t)\rangle -\langle x(t) \rangle^2}{2 t}\right]\,,\label{eq:DeffGen}
\end{align}
where $\langle x(t)\rangle$ and $\langle x^2(t)\rangle$ are respectively the conditional first moment and second moment given by
\begin{align}
  \langle x(t)\rangle &= \frac{\int_{-\infty}^\infty dx\, x\, p(x,t)}{S(t)}\,,\qquad \langle x^2(t)\rangle = \frac{\int_{-\infty}^\infty dx\, x^2\, p(x,t)}{S(t)}\,,\label{eq:sm}
\end{align}
where $S(t)\coloneqq S(t\,|x_0=0)$ is the survival probability of the particle at time $t$ given that it started from $x_0=0$, which was studied in \cite{GPBD21}, and $p(x,t)$ is the unconditional probability distribution function of the position $x$ of the Brownian motion in the presence of the traps at time $t$. The propagator $p(x,t)$ satisfies the forward Fokker-Planck equation
\begin{align}
  \partial_t p(x,t) = D \partial_{xx} p(x,t)- \beta \sum_{m=-\infty}^{\infty} \delta\left(x-\frac{L}{2}-m\,L\right)p(x,t)\,,\label{eq:fp}
\end{align}
where the last term in the right-hand side accounts for the partially absorbing traps of intensity $\beta$ separated by a distance $L$ and evenly spaced around the initial position $x_0=0$. The differential equation (\ref{eq:fp}) must be solved with the initial condition $p(x,t=0)=\delta(x)$ as the particle starts from the origin $x_0=0$. This is a rather difficult task since the initial condition breaks the periodic symmetry $x\to x+L$. Nevertheless, by establishing a connection with a similar albeit different problem, which concerns the winding statistics of a Brownian motion on a ring \cite{Kundu(2014)}, we obtain an exact closed-form expression for the effective coefficient in (\ref{eq:DeffGen}). Furthermore, we provide a rejection-free algorithm, based on an effective Langevin equation, to generate surviving particles, which is particularly useful for numerical purposes. We show that the point absorbers induce an effective repulsive potential on the surviving particles. For pedagogical purposes, we first deal with the case of a finite number of traps, after which we solve the case of an infinite number of traps, which turns out to be quite different and non-trivial. Finally, we generalise our results to other trapping environments.

\subsection{Presentation of our results}
It is useful to summarise our main results. Let us first present the effective diffusion coefficient in the presence of a finite number of traps $n$
 and then discuss the case of an infinite number of point absorbers.
\subsubsection{Finite number of traps}
We find that as long as $n$ is finite, the effective diffusion coefficient (\ref{eq:DeffGen}) is simply
\begin{align}
  D_\text{eff} = 2D\,,\label{eq:nfinite}
\end{align}
independently of the distance $L$ and of the trapping intensity $\beta$. This result can be easily understood. The surviving particles typically diffuse away from the traps towards $x\to \pm \infty$ (see the caption of figure \ref{fig:model}). The probability density function can therefore be approximated by a superposition of two Gaussian distributions traveling in opposite directions away from the origin. And, consequently, the effective diffusion coefficient is twice the original one. 

\subsubsection{Infinite number of traps}
\begin{figure}
  \begin{center}
    \includegraphics[width=0.5\textwidth]{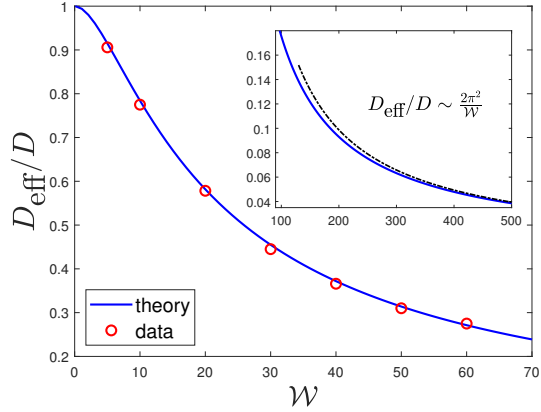}
    \caption{Normalised effective diffusion coefficient $D_\text{eff}/D$ as a function of the Sherwood number $\mathcal{W}=\beta L/D$ for a diffusive particle, starting from the origin, conditioned to survive in the presence of infinitely many partially absorbing traps with intensity $\beta$, periodically distributed with a period $L$ and evenly spaced around the origin. The blue line is the theoretical prediction in (\ref{eq:Deff}), whereas data points are obtained by simulating $3\cdot 10^8$ Gaussian random walks up to step $2\cdot 10^5$. The time increment is set to $\tau=0.001$, the variance is set equal to $\sigma^2\coloneqq 2D\tau$ with $D=1$ and the point absorbers become trapping intervals of length $\beta\tau$ separated by a distance $L=10$. }
    \label{fig:Deff}
  \end{center}
\end{figure}

The situation is drastically different when there is an infinite number of point absorbers $n\to \infty$ as the surviving particles cannot escape the traps by diffusing away to $x\to \pm \infty$. Instead, they typically stay in-between the point absorbers and the effective diffusion coefficient (\ref{eq:DeffGen}) is given by
\begin{align}
  D_\text{eff} = D \mathcal{F}\left(\mathcal{W}=\frac{\beta L}{D}\right)\,,\label{eq:Deffinf}
\end{align}
with the scaling function $\mathcal{F}\left(\mathcal{W}\right)$ defined as
\begin{align}
 \mathcal{F}(\mathcal{W}) =\left[\sin ^2\left(\frac{\sqrt{\mathcal{G}(\mathcal{W})}}{2}\right)\left(\frac{2}{\mathcal{G}(\mathcal{W})}+\frac{2}{\mathcal{W}}+\frac{\mathcal{W}}{2\,\mathcal{G}(\mathcal{W})}\right)\right]^{-1}\,,\label{eq:Deff}
\end{align}
 where $\mathcal{W}$ is the Sherwood number and the scaling function $\mathcal{G}(\mathcal{W})$ is given implicitly as the first zero of the transcendental equation
 \begin{align}
  \cot\left(\frac{\sqrt{\mathcal{G}(\mathcal{W})}}{2}\right) = \frac{2\sqrt{\mathcal{G}(\mathcal{W})}}{\mathcal{W}}\,.\label{eq:gu}
\end{align}
The Sherwood number $\mathcal{W}$ is a dimensionless number in fluid mechanics that represents the ratio of convective mass transfer over diffusive mass transport \cite{Sherwood}. Interestingly the scaling function defined in (\ref{eq:gu}) is the same as the one found in \cite{GPBD21} that governs the decay rate of the survival probability. A plot of the function $\mathcal{G}(\mathcal{W})$ is given in figure 5 in \cite{GPBD21}. The asymptotic behaviors of $\mathcal{F}(\mathcal{W})$ are given by
\begin{align}
  \mathcal{F}(\mathcal{W}) \sim \left\{\begin{array}{ll}
    1\,,\qquad&\mathcal{W} \to 0\,,\\[1em]
   \dfrac{2\pi^2}{\mathcal{W}} \,,\qquad&\mathcal{W} \to \infty\,.
  \end{array}\right.
\end{align}
A plot of the scaling function (\ref{eq:Deff}) is shown in figure \ref{fig:Deff}. As one can see, the effective diffusion coefficient is always less that one, contrary to the case of a finite number of traps. As the surviving particles cannot escape to $x\to \pm \infty$, they tend to stay in-between traps, hence having a smaller diffusion coefficient than the original one. Similar results can be found in \cite{Powles1992, Tanner1978} for diffusion impeded by semi-permeable barriers or heat flow in the presence of planar slabs with a finite contact resistence at the interfaces. 

\begin{figure}[htbp]
  \begin{center}
    \includegraphics[width=0.6\textwidth]{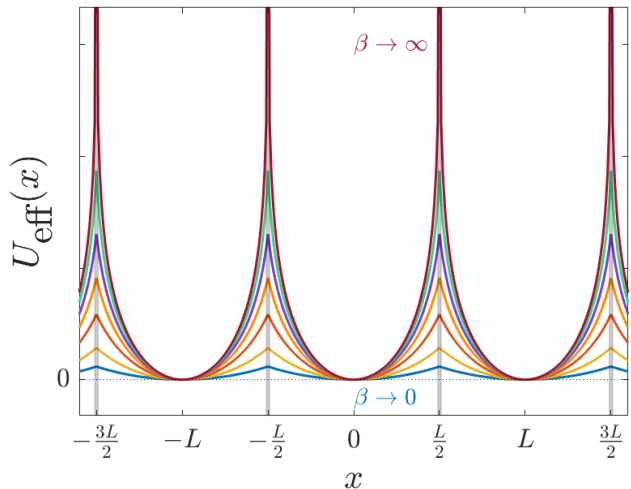}
    \caption{Effective potential $U_\text{eff}(x)$ induced by the periodically distributed partially absorbing traps on the surviving particles as a function of $x$ for several trap intensities $\beta$ [see expression (\ref{eq:Ueff})]. As the trap intensity tends to infinity $\beta\to \infty$, the potential diverges at the location of the traps.}
    \label{fig:EffPot}
  \end{center}
\end{figure}

Finally, we obtain a rejection-free algorithm based on an effective Langevin equation in order to generate only surviving trajectories. We find that the effective Langevin equation that governs the evolution of the surviving particles is given by
\begin{align}
   \dot x_s(t) = \sqrt{2D}\,\eta(t) - \partial_{x_s} U_\text{eff}(x_s)\,,\label{eq:effLf}
\end{align}
where the subscript $s$ in $x_s(t)$ refers to ``surviving'' trajectories and the effective repulsive potential $U_\text{eff}(x)$ induced by the infinite number of traps is defined as 
\begin{align}
  U_\text{eff}(x) = - 2D \ln\left[\cos\left(\frac{x\sqrt{\mathcal{G}\left(\mathcal{W}\right)}}{L}\right)\right]\,,\qquad -\frac{L}{2}\leq x\leq \frac{L}{2}\,,\label{eq:Ueff}
\end{align}
and is $L$-periodic $U_\text{eff}(x)=U_\text{eff}(x+L)$. We recall that $\mathcal{G}(\mathcal{W}=\beta L/D)$ is given in (\ref{eq:gu}). The effective potential in (\ref{eq:Ueff}) takes a remarkably simple form with cusps located at the positions of the point absorbers (see figure \ref{fig:EffPot}). The effective Langevin equation in (\ref{eq:effLf}) can be discretised over small time increments to  generate surviving trajectories in an efficient manner.

The rest of the paper is organized as follows. In section~\ref{Sec: finite}, we focus in detail on the cases of a single trap and a finite number $n$ of point absorbers symmetrically distributed around the initial condition. In section~\ref{Sec: infinite}, we discuss the case of an infinite number of traps. In section~\ref{Sec: various}, we generalize the previous basic environments by considering asymmetric geometries or starting positions, that can induce also a permanent drift in addition to the effective diffusion coefficient. Finally, we give  a concluding summary and some perspectives.

\section{Finite number of traps}
\label{Sec: finite}
\subsection{Single trap}\label{subsec:single}
We start by the exactly solvable case of a single trap $n=1$ located at $x_1=0$. The propagator $p(x,t)$, which is the probability density function of the position $x$ at time $t$, satisfies the Fokker-Planck equation
\begin{align}
  \partial_t p(x,t) = D \partial_{xx} p(x,t) -\beta \delta(x) p(x,t)\,,\label{eq:fp1}
\end{align}
with the initial condition $p(x,0)=\delta(x-x_0)$, where $x_0$ is the initial position of the particle. Without loss of generality, we assume that the initial position is positive $x_0\geq 0$ and set it to zero $x_0=0$ afterwards. We will solve the Fokker-Planck equation (\ref{eq:fp1}) to obtain the expression of the propagator $p(x,t)$. Then, we will insert it into the expressions (\ref{eq:sm}) and (\ref{eq:DeffGen}) to obtain the effective diffusion coefficient. 

In order to solve the Fokker-Planck equation in (\ref{eq:fp1}), we turn to the Laplace domain in which it reads
\begin{align}
  s \tilde p(x,s) -\delta(x-x_0) = D \partial_{xx}\tilde p(x,s) -\beta \delta(x)\tilde  p(x,s)\,,\label{eq:fp1s}
\end{align}
where $\tilde  p(x,s)  \coloneqq \int_0^\infty dt e^{-st}p(x,t)$. The equation (\ref{eq:fp1s}) can be solved on the three separate intervals: $x<0$, $0<x<x_0$, and $x>x_0$. The general solution is
\begin{align}
  \tilde p(x,s) = \left\{\begin{array}{ll}
    A(s) e^{x \sqrt{\frac{s}{D}}}\,, \qquad& x< 0\,,\\
    B(s) e^{x \sqrt{\frac{s}{D}}} + C(s)  e^{-x \sqrt{\frac{s}{D}}}\,, \qquad& 0<x<x_0\,,\\
    E(s) e^{-x \sqrt{\frac{s}{D}}}\,, \qquad& x>x_0\,,
  \end{array}\right.\label{eq:gens}
\end{align}
where $A(s)$, $B(s)$, $C(s)$ and $E(s)$ are integration constants, i.e.~independent of $x$. By imposing the continuity of the solution and matching the derivative by integrating (\ref{eq:fp1s}) over an infinitesimal interval around $x$ and $x_0$, we obtain the integration constants
\begin{align}
  A(s) &= \frac{e^{-\sqrt{\frac{s}{D}} x_0}}{2\sqrt{s D}+\beta }\,,\qquad B(s) = \frac{e^{-\sqrt{\frac{s}{D}} x_0}}{2\sqrt{s D}}\,,\qquad C(s) = -\frac{\beta e^{-\sqrt{\frac{s}{D}} x_0}}{4Ds+2\sqrt{s D}\beta}\,,\nonumber\\
      E(s) &= \frac{1}{2\sqrt{sD}+\beta}\left[ \exp\left(\sqrt{\frac{s}{D}}x_0\right)+ \frac{\beta}{\sqrt{sD}}\sinh\left(\sqrt{\frac{s}{D}}x_0\right)\right]\,.\label{eq:ABCE}
\end{align}

From the explicit expression of the propagator in (\ref{eq:gens}), we obtain the survival probability which is needed to normalise the conditional first and second moment in (\ref{eq:sm}).
The Laplace transform of the survival probability $S(t)$ is given by
\begin{align}
  \tilde S(s) =\int_0^\infty dt S(t)e^{-st} = \int_{-\infty}^{\infty} dx \,\tilde p(x,s) = \frac{1}{s}-\frac{\beta}{s}\frac{e^{-\sqrt{\frac{s}{D}} x_0}}{2\sqrt{sD}+\beta}\,,\label{eq:Ss}
\end{align}
which can be inverted exactly and yields the scaling form 
\begin{align}
  S(t) = \mathcal{S}\left(y=\frac{x_0}{\sqrt{Dt}},\,b=\beta \sqrt{\frac{t}{D}}\right)\,,\label{eq:Sb}
\end{align}
where the scaling function $\mathcal{S}$ is given by
\begin{align}
\mathcal{S}(y,b)=  \text{erf}\left(\frac{y}{\sqrt{4\pi}}\right) + y\int_0^1du \frac{1}{\sqrt{4\pi u^3}}\text{erfc}\left(\frac{b}{2} \sqrt{1-u}\right) e^{-\frac{y^2}{4u}+\frac{b^2(1-u)}{4}}\,.\label{eq:f}
\end{align}
The expression (\ref{eq:f}) has a simple interpretation: it is given by the survival probability in the presence of a fully absorbing trap at the origin plus a correction, which accounts for the partial absorption. One can check that the correction vanishes when $b\to \infty$, corresponding to an absorbing barrier, and that we recover $\mathcal{S}(y,0)=1$ when $b=0$, which stands for the absence of a trap. 
% When $y=0$, we find
% \begin{align}
%   \mathcal{S}(y=0,b) =e^{\frac{b^2}{4}} \text{erfc}\left(\frac{b}{2}\right)\,.\label{eq:Sx00}
% \end{align}
% The asymptotic behavior of $S(x_0,t)$ for large $t$ is given by
% \begin{align}
%   S(x_0,t) \sim \left(\frac{2\sqrt{D}}{\beta \sqrt{\pi}} + \frac{x_0}{\sqrt{\pi D}}\right)\frac{1}{\sqrt{t}}\,,\quad t \to \infty\,.\label{eq:Sa}
% \end{align}
% The uneffective average position $\langle x(t)\rangle$ is
% \begin{align}
%   \langle x(t)\rangle = \int_{-\infty}^{\infty} dx  x p(x,t) = x_0\,,\label{eq:x0}
% \end{align}
% The effective average position $\langle x(t)\rangle$ is
% \begin{align}
%   \langle x(t)\rangle = \frac{\langle x(t)\rangle}{S(x_0,t)} = \frac{x_0}{S(x_0,t)}\,,\label{eq:unc}
% \end{align}
% where $S(x_0,t)$ is given in (\ref{eq:Sb}).
% We find that the asymptotic behavior of  $  \langle x(t)\rangle $ is 
% \begin{align}
%  \langle x(t)\rangle \sim  \frac{x_0\beta \sqrt{\pi Dt}}{2 D+\beta x_0}\,,\quad t\to \infty\,.
% \end{align}

 The Laplace transform of the second moment in (\ref{eq:sm}) can also be extracted from the probability distribution (\ref{eq:gens}) and we get
\begin{align}
\int_{-\infty}^\infty dx \,x^2\,\tilde p(x,s) = \frac{2D}{s^2}-\frac{2D\beta e^{-\sqrt{\frac{s}{D}}x_0}}{s^2 (2\sqrt{s D}+\beta)}\,.\label{eq:2s}
\end{align}
The Laplace transform in (\ref{eq:2s}) can be inverted exactly and provides the scaling form for the unconditional second moment
\begin{align}
\int_{-\infty}^\infty dx \,x^2\, p(x,t) =  x_0^2 +  2Dt\,\mathcal{G}\left(y=\frac{x_0}{\sqrt{Dt}},\,b=\beta \sqrt{\frac{t}{D}}\right)\,,\label{eq:x2}
\end{align}
where the scaling function $\mathcal{G}$ is given by
\begin{align}
\mathcal{G}(y,b) = 1 -  \int_0^1 du \left[1-e^{\frac{b^2 u }{4 }} \text{erfc}\left(\frac{b}{2}\sqrt{u}\right)\right] \text{erfc}\left(\frac{y}{1 \sqrt{1-u}}\right)\,.\label{eq:g}
\end{align}
Similarly to the survival probability, the unconditional second moment can be seen as the second moment in the presence of a fully absorbing trap plus a correction, which accounts for the partial absorption. Again, one can check that the correction vanishes when $b\to \infty$, and that we recover $\mathcal{G}(y,0)=1$ when $b=0$. 

\begin{figure}[htbp]
  \begin{center}
    \includegraphics[width=0.6\textwidth]{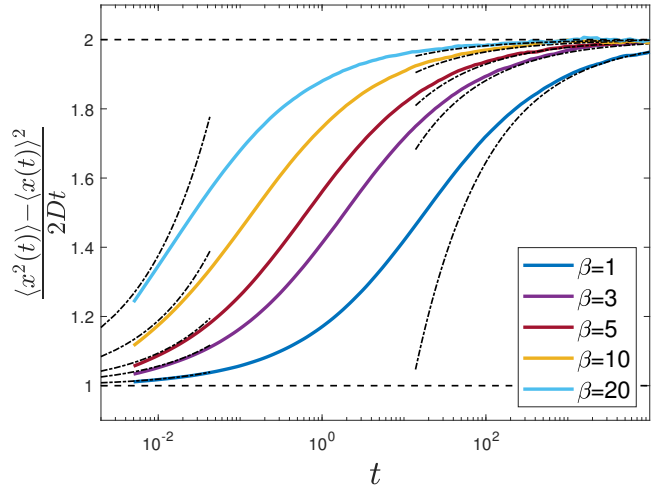}
    \caption{Time evolution of the normalised mean-square displacement per unit time for a diffusive particle, starting from the origin, which survives in the presence of a single trap located at the origin. Dash-dotted lines represent the theoretically predicted asymptotic behaviours at short and long times, see \eqref{eq:as}. Colored lines depict numerical results for $ 10^8$ Gaussian random walks evolved up to step $10^7$, with time increment $\tau=0.001$, $D=1$, variance $2\tau$ and a trapping interval at the origin of length $\beta\tau$.}
    \label{fig:1TrapOrigin}
  \end{center}
\end{figure}

The unconditional first moment can be obtained similarly from the probability distribution, or alternatively by using the martingale property of the process which directly gives 
\begin{align}
\int_{-\infty}^\infty dx \,x\, p(x,t)=x_0\,.\label{eq:fmom}
\end{align}
Inserting the expressions of the unconditional first and second moments (\ref{eq:fmom})-(\ref{eq:x2}), and the survival probability (\ref{eq:Sb}), we find that the term in brackets in (\ref{eq:DeffGen}) becomes
\begin{align}
  \frac{\langle x^2(t)\rangle -\langle x(t) \rangle^2}{2 t} = D\,\frac{\mathcal{G}\left(y,b\right)}{\mathcal{S}\left(y,b\right)}\,,\label{eq:brack}
\end{align}
where $y=x_0/\sqrt{Dt}$ and $b=\beta \sqrt{t/D}$. By setting $x_0=0$ and taking the limit $t\to \infty$ in (\ref{eq:brack}), one immediately finds the result (\ref{eq:nfinite}) announced in the introduction.
%  When $y=0$, it gives
%  \begin{align}
%    g(y=0,b)= 4\left[\frac{e^{\frac{b^2}{4}}}{b^2}\text{erfc}\left(\frac{b}{2}\right)+\frac{1}{b\sqrt{\pi}}-\frac{1}{b^2}\right]\,.\label{eq:gy0}
%  \end{align}
%\begin{align}
%  \langle x(t)^2\rangle = \frac{ \langle \tilde x(t)^2\rangle}{S(x_0=0,t)} = \frac{8D^2}{\beta^2} + \frac{8 e^{-\frac{t \beta^2}{4D}}(\beta\sqrt{D^3 t} -D^2 \sqrt{\pi})}{\sqrt{\pi}\beta^2 \text{erfc}\left(\frac{1}{2}\beta \sqrt{\frac{t}{D}}\right)}\,,\label{eq:x2c}
%\end{align}
%whose asymptotic behavior is given by
%\begin{align}
%  \langle x(t)^2\rangle = \left\{ \begin{array}{ll}
%    2D t + \frac{2\sqrt{D}\beta t^{\frac{3}{2}}}{3 \sqrt{\pi}}\,,& t\to 0\,,\\
%    4 D t - \frac{4 (D^{3/2}\sqrt{\pi})\sqrt{t}}{\beta}\,,& t\to \infty\,.
%  \end{array}\right.\label{eq:x2cl}
%\end{align}
The expression in (\ref{eq:brack}) contains the full time dependence and, in particular, describes the transient regime of anomalous diffusion. It takes a rather simple form when $x_0=0$, i.e.~when the particle starts exactly in the partially absorbing trap, which reads
\begin{align}
   \frac{\langle x^2(t)\rangle -\langle x(t) \rangle^2}{2 t} = D\,\frac{\mathcal{G}\left(0,b\right)}{\mathcal{S}\left(0,b\right)}= \frac{4D}{e^{\frac{b^2}{4}} \text{erfc}\left(\frac{b}{2}\right)} \left[\frac{e^{\frac{b^2}{4}}}{b^2}\text{erfc}\left(\frac{b}{2}\right)+\frac{1}{b\sqrt{\pi}}-\frac{1}{b^2}\right]\,,\label{eq:x2c}
\end{align}
where $b=\beta \sqrt{t/D}$.
In this case, we therefore find that the asymptotic behavior is given by
\begin{align}
  \frac{\langle x^2(t)\rangle -\langle x(t) \rangle^2}{2D t}  \sim \left\{\begin{array}{ll}
    1+\frac{b}{3\sqrt{\pi}}\,, \qquad& b\to 0\,,\\
    2-\frac{2\sqrt{\pi}}{b}\,, \qquad& b\to \infty\,.
  \end{array}\right. \label{eq:as}
\end{align}
The time evolution of the normalised mean-square displacement per unit time in \eqref{eq:x2c} is shown in figure~\ref{fig:1TrapOrigin}. Note that, as the particle is initially located in the partially absorbing trap, the initial regime of pure diffusion where the particle should not ``see'' the trap is not present, and the transient regime of anomalous diffusion directly takes place. The typical crossover timescale to fast normal diffusion is of the order of $D/\beta^2$, as established in \eqref{eq:as}. 

As a final comment, we point out the commutation issue in the limits $\beta\to 0$ and $t\to \infty$: when $\beta=0$ one clearly expects that the effective diffusion coefficient coincides with the original one, since particles are definitively free. When $\beta\to 0$, instead, they can eventually be absorbed, as a consequence the ratio $D_\text{eff}/D$ still tends to $2$.

\subsection{A finite number of traps}\label{sec:arbFinite}
In the previous section, we showed that for a trapping environment with a single trap and symmetric initial condition, the effective diffusion coefficient for the surviving particles is equal to twice the original one.
In this section, we extend this result to a symmetric trapping environment with $n$ traps (see figure \ref{fig:model}). To do so, one could perform a similar computation as in the previous section with $n$ point absorbers. However, this calculation becomes quite cumbersome for large values of $n$. In \ref{sec:two}, we perform the full computation for $n=2$ traps and show that the effective diffusion coefficient is equal to $2D$, as stated in (\ref{eq:nfinite}). We claim that this result holds for any finite number of point absorbers. This claim is based on the following argument. In the long time limit, the probability distribution will take a diffusive scaling form $p(x,t)=\frac{1}{\sqrt{Dt}} \mathcal{P}(x/\sqrt{Dt})$. Upon inserting this expression in the Fokker-Planck equation, one sees that, in the limit $t\to \infty$, the multiple traps effectively become a single trap located at the origin with intensity $n\beta$. In other words, the length scale of the trapping region becomes irrelevant in the long time limit. Hence, as the result for the single point absorber does not depend on the trap intensity, it is still valid for a finite number $n$ of traps.

\section{Infinitely many traps}\label{Sec: infinite}
\subsection{Effective diffusion coefficient}
In the previous section, we saw that, in the case of a finite number of traps, the effective diffusion coefficient of the surviving particles is systematically equal to twice the original diffusion coefficient. In this section, we study the case of infinitely many point absorbers located at $\pm(L/2+jL)$ with $j\in \mathbb{N}_0$ (see figure \ref{fig:model}). One can already notice that the argument used for a finite number of traps breaks down as the effective single trap intensity tends to infinity. It is therefore necessary to find a different way to compute the effective diffusion coefficient, without solving the Fokker-Planck equation in (\ref{eq:fp}), which seems intractable for infinitely many traps. 
\begin{figure}
  \begin{center}
  \includegraphics[width=0.22\textwidth]{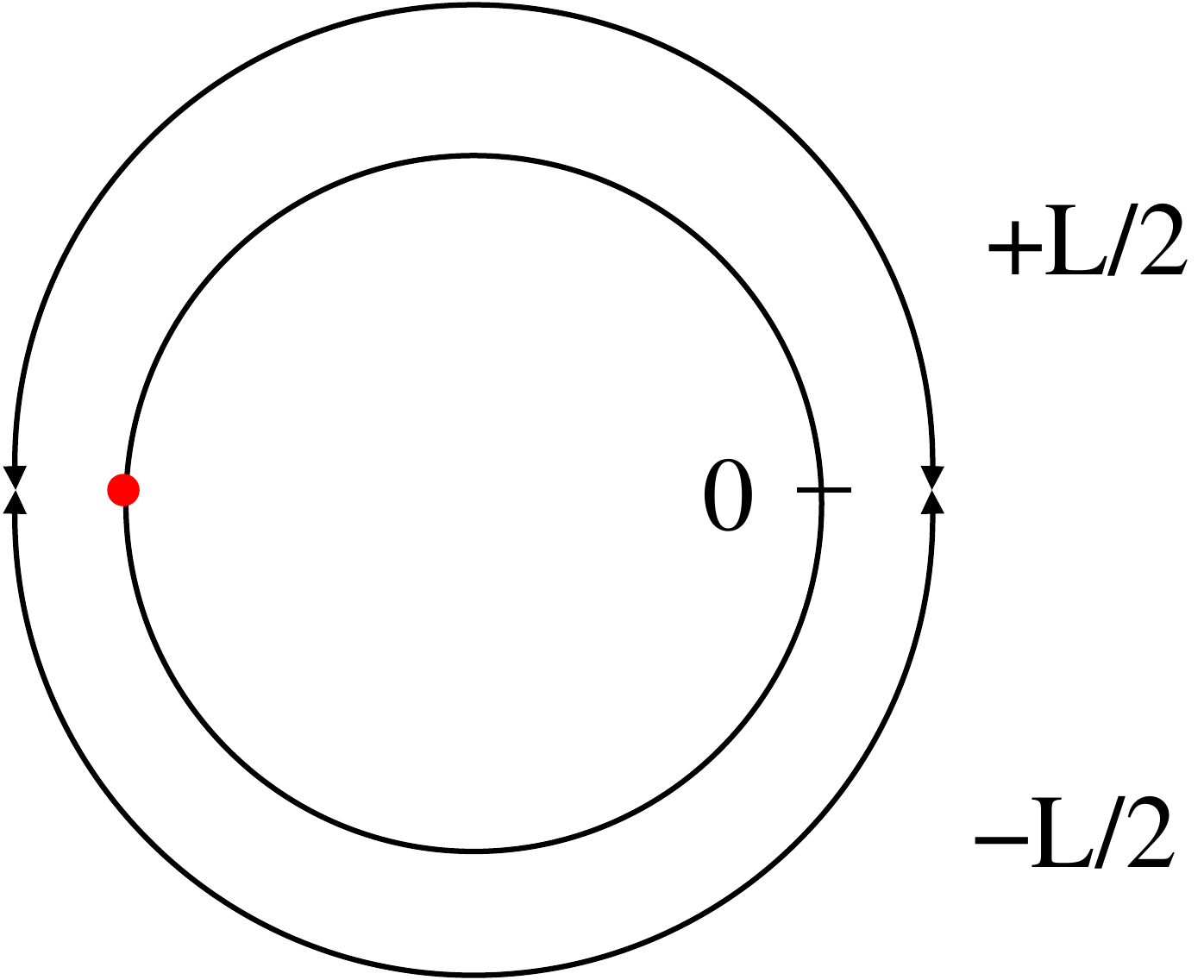}\hspace{2em}
    \includegraphics[width=0.4\textwidth]{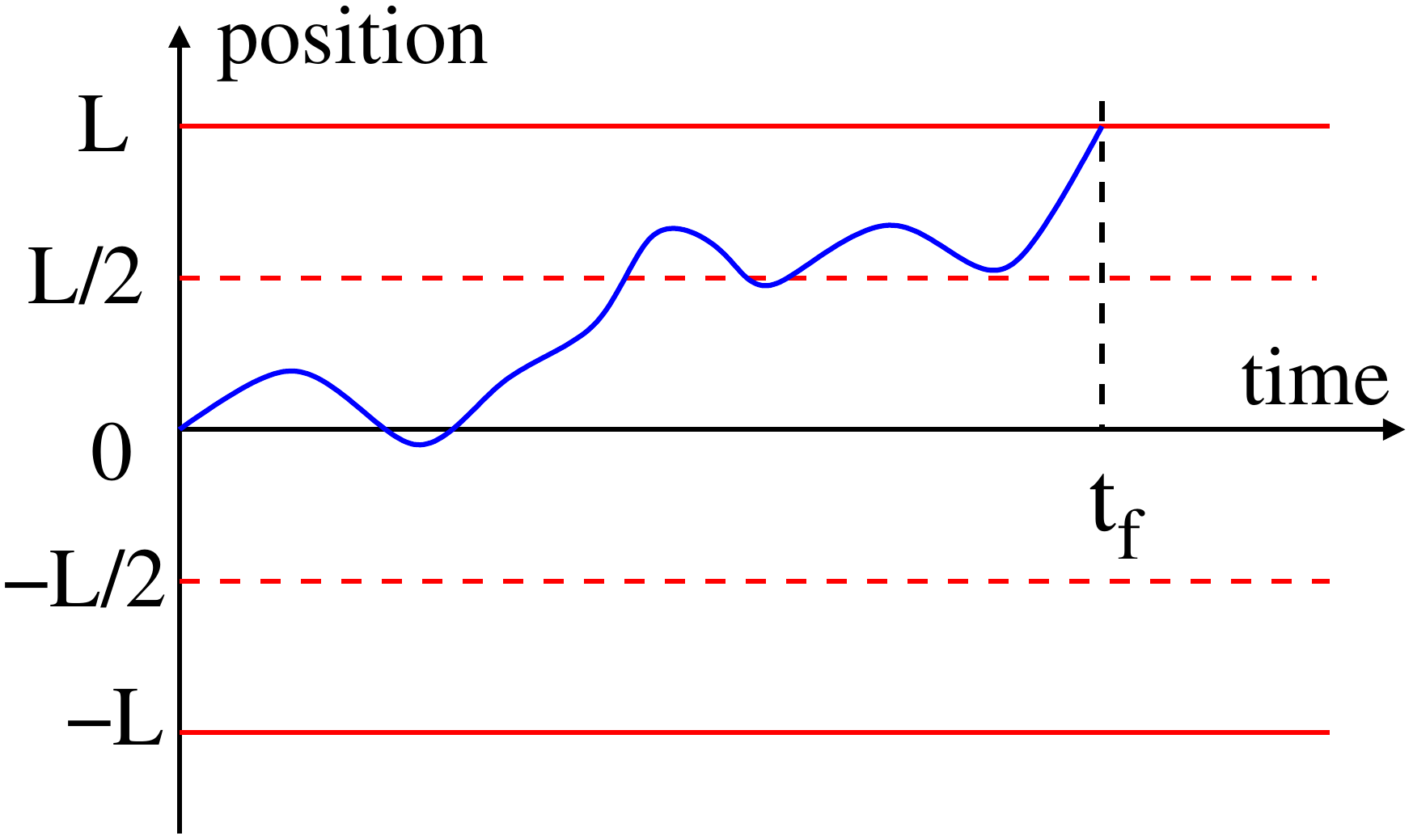}
    \caption{\textbf{Left panel:} Mapping of the periodic trapping environment in figure \ref{fig:model} to a circle of perimeter $L$ with a single partially absorbing trap (red dot) located at arc length $\pm L/2$. As time grows, the surviving particle crosses several traps in figure \ref{fig:model}, which corresponds to several turns around the circle. More precisely, when the particle is exactly in the middle of two point absorbers and propagates to the middle of one of the two neighboring regions in figure \ref{fig:model}, it makes a full turn around the circle and the process starts anew.  Depending on which direction it goes, the particle turns either in a clockwise or anti-clockwise fashion. We track the position on the $x$-axis in figure \ref{fig:model} by counting the net number of full revolutions around the circle. \textbf{Right panel:} Each complete turn of the circle in a time $t_f$ in the left panel is associated with a first-exit event of a box $[-L,L]$ at time $t_f$ with two partially absorbing traps located at $\pm L/2$. A full clockwise turn corresponds to exiting the box at $+L$, whereas a full anti-clockwise turn corresponds to exiting the box at $-L$. The distribution of the exit time $t_f$, either at $+L$ or $-L$, is denoted by $f(t_f)$. The survival probability at time $\tau$, where the particle does not exit the box nor gets trapped until time $\tau$, is denoted by $S_\text{box}(\tau)$.}
    \label{fig:mapping}
  \end{center}
\end{figure}
We found a way to obtain the effective diffusion coefficient by relying on an idea developed in \cite{Kundu(2014)} on a seemingly different problem concerning the winding statistics of a Brownian motion on a ring. The main idea is as follows. We map the periodic structure of the environment to a circle of perimeter $L$ with a single partially absorbing trap located at arc length $\pm L/2$ (see figure \ref{fig:mapping}). As time increases, the particle turns around the circle either in a clockwise or anti-clockwise fashion until it is absorbed by one of the traps. The presence of the point absorbers makes this problem inherently different from \cite{Kundu(2014)}. Using a similar notation as in the original paper \cite{Kundu(2014)}, we denote by $m_+(t)$ the total number of complete counter-clockwise turns and $m_-(t)$ the total number of complete clockwise turns after time $t$. Furthermore, we denote by $m(t)=m_+(t)+m_-(t)$ the absolute number of complete turns and by $k(t)=m_+(t)-m_-(t)$ the net number of full revolutions after time $t$. Then we introduce the probability distribution $R(k,t)$ of the net number of complete turns $k$ after time $t$:
\begin{align}
  R(k,t) = \text{Prob.}\left(k(t) = k\right)\,.\label{eq:Rkt}
\end{align}
 Due to the presence of the traps, the distribution $R(k,t)$ is not normalised to unity and is in fact normalised to the survival probability $S(t)\coloneqq S(t|x_0=0)$ with the initial position set equal to $x_0=0$. The normalisation reads
\begin{align}
  S(t) = \sum_{k=-\infty}^\infty R(k,t)\,.\label{eq:SrelRk}
\end{align}
The net number of complete laps is directly related to the position of the particle on the real line by
\begin{align}
  k(t) = \left\lfloor \frac{x(t)}{L} \right\rfloor \,,\label{eq:kxrel}
\end{align}
where $\lfloor a\rfloor$ denotes the floor operation returning the highest integer lower or equal to $a$. In the long time limit, we expect that the unconditional second moment on the real line behaves as
\begin{align}
  \int_{-\infty}^\infty dx\,x^2\, p(x,t) \approx L^2\,\sum_{k=-\infty}^\infty k^2\,R(k,t)\,,\qquad t\to \infty\,.\label{eq:x2k2}
\end{align}
In fact, in this limit, one can neglect the dynamics inside a box of length $2L$ and record the trajectory on a large scale as jumps between the centres of these intervals, numerated by the net winding number $k$.
By symmetry of the trapping environment, it is clear that $\langle x(t)\rangle =0$, and inserting the expressions (\ref{eq:SrelRk})-(\ref{eq:x2k2}) into (\ref{eq:sm}), we find that the effective diffusion coefficient (\ref{eq:DeffGen}) can be written as
\begin{align}
  D_{\text{eff}} = \lim_{t\to \infty}\frac{L^2\,\sum_{k=-\infty}^\infty k^2\,R(k,t)}{2\,t\,\sum_{k=-\infty}^\infty R(k,t)}\,.\label{eq:Deffint}
\end{align}
We will now show how to compute $R(k,t)$ using a similar approach as in \cite{Kundu(2014)}.

In order to study the distribution $R(k,t)$, it is first convenient to analyse a more general quantity, $P(m,k,t)$, which is the joint distribution of the absolute number of full revolutions $m$ and the net number of complete turns $k$ after time $t$:
\begin{align}
  P(m,k,t) = \text{Prob.}\left(m(t)=m,k(t)=k\right)\,.\label{eq:defpmkt}
\end{align}
 From this joint distribution, we obtain $R(k,t)$ by summing over all values of $m$, i.e.
\begin{align}
  R(k,t) = \sum_{j=0}^\infty P(2j+|k|,k,t)\,,\label{eq:RP}
\end{align}  
where we used the fact that the absolute number of laps $m$ can be written as a sum of an even number $2j$ of full turns plus the absolute value of the net number of complete revolutions $|k|$.  We first count the total number of turns $m$ completed before the particle gets absorbed by a trap, say after a time $t$, and the corresponding durations $\{\tau_1,\ldots,\tau_m\}$ between consecutive windings. Then we denote by $\tau_{\text{last}}=t-\sum_{i=1}^m \tau_i$ the duration of the last incomplete turn. The number of laps $m$ and the durations $\{\tau_1,\ldots,\tau_m,\tau_{\text{last}}\}$ are random variables whose joint probability distribution is given by
\begin{equation}
\mathcal{P}(m, \{\tau_1,\tau_2,\dots,\tau_m,\tau_{last}\}, t)=f(\tau_1)f(\tau_2)\dots f(\tau_m)S_\text{box}(\tau_{\text{last}})\delta \left(t-\tau_{last}-\sum_{i=1}^m\tau_i\right)\,,\label{eq:Pturn}
\end{equation}
 where $f(\tau)$ is the first-exit probability from a box $[-L,L]$, with centered initial position and partially absorbing traps of strength $\beta$ located at $\pm L/2$, whereas $S_\text{box}(\tau)$ is the survival probability in the box, i.e. the probability that the particle did not exit the box and did not get trapped (see figure \ref{fig:mapping}). The first $m$ terms in the right-hand side of \eqref{eq:Pturn} refer to the first $m$ laps, the factor $S_\text{box}(\tau_{\text{last}})$ refers to the time spent after the last complete turn, and the final Dirac delta term imposes the constraint that $t=
\tau_{last}+\sum_{i=1}^m\tau_i$. The marginal distribution of the total number of complete turns $\mathcal{P}(m,t)$ can be obtained by integrating over all possible durations $\{\tau_1,\ldots,\tau_m,\tau_{\text{last}}\}$:
\begin{align}
  P(m,t)=\int_0^\infty d\tau_1\ldots d\tau_m d\tau_{\text{last}}f(\tau_1)f(\tau_2)\dots f(\tau_m)S_\text{box}(\tau_{\text{last}})\delta \left(t-\tau_{last}-\sum_{i=1}^m\tau_i\right)\,.\label{eq:Pturn2}
\end{align}
From the distribution of the absolute number of complete turns in (\ref{eq:Pturn2}), we obtain the joint distribution $P(m,k,t)$ with the net number of complete laps $k$ using the following argument. Given that the particle made $m$ absolute number of complete turns, the particle will have made $m_+$ counter-clockwise revolutions with probability $(1/2)^{m_+}$ and $m_-$ clockwise turns with probability $(1/2)^{m_-}$, as when the particle exits the box, it is located in the middle of the neighbouring interval (see figure \ref{fig:mapping}). By using the relations $m_+=(m+k)/2$ and $m_-=(m-k)/2$, and summing over all possible combinations of $m$ and $k$, we find that%ways of having $m$ absolute and $k$ net number of windings, we find that 
\begin{align}
  P(m,k,t) = P(k|m,t) P(m,t) = \binom{m}{\frac{m+k}{2}}\left(\frac{1}{2}\right)^m P(m,t)\,,\label{eq:relPmkt}
\end{align}
where the binomial coefficient counts the number of ways to make $m_+=(m+k)/2$ counter-clockwise turns among $m$ turns and the factor $\left(\frac{1}{2}\right)^m$ is the probability of each configuration. 
Inserting the expression (\ref{eq:Pturn2}) into (\ref{eq:relPmkt}) and noting that the convolution structure over time of this integral is well suited to a Laplace transform, we find that in Laplace domain $P(m,k,t)$ reads
\begin{align}
 \tilde P(m,k,s)= \int_0^\infty dt e^{-st} P(m,k,t) = \binom{m}{\frac{m+k}{2}}\left(\frac{1}{2}\right)^m\tilde f^m (s)\tilde S_\text{box}(s)\,,\label{eq:tildPms}
\end{align}
where $\tilde f (s)$ and $\tilde S_\text{box}(s)$ are the Laplace transforms of $f(t)$ and $S_\text{box}(t)$ respectively. Thus, taking a Laplace transform of the expression (\ref{eq:RP}), we find that in Laplace domain the probability $R(k,t)$ for a net winding number $k$ in a time $t$ is given by
\begin{equation}
\tilde{R}(k,s)=\sum_{m=0}^{\infty}\binom{2m+|k|}{m}\left(\frac{\tilde{f}(s)}{2}\right)^{2m+|k|}\tilde{S}_\text{box}(s)\,.
\end{equation}
More explicitly, using the identity \cite{wiki}
$$\sum_{m=0}^{\infty}\binom{2m+|k|}{m} z^m= \frac1{\sqrt{1-4z}}\left(\frac{1-\sqrt{1-4z}}{2z}\right)^{|k|}\,,\qquad |z|<\frac 1 4\,,$$
we get
\begin{align}
  \tilde R(k,s) = \frac{\tilde S_\text{box}(s)}{\sqrt{1-\tilde f^2(s)}}\left(\frac{1-\sqrt{1-\tilde f^2(s)}}{\tilde f(s)}\right)^{|k|}\,.\label{eq:tR}
\end{align}
The expressions of the first-passage distribution $\tilde f(s)$ and the survival probability $\tilde S_\text{box}(s)$ can be computed straightforwardly from a Fokker-Planck equation (see \ref{app:InfBox}), and they are found to be given by
\begin{align}
 \tilde f(s) &= \left[\frac{\beta}{2 \sqrt{sD} }  \sinh    \left(L \sqrt{\frac{s}{D}}\right)+\cosh \left(L    \sqrt{\frac{s}{D}}\right)\right] ^{-1}\,,\label{eq:fs}\\
 \tilde S_\text{box}(s) &= \frac{1}{s}\left[1-\frac{1}{\frac{\beta}{2 \sqrt{sD} }  \sinh
   \left(L \sqrt{\frac{s}{D}}\right)+\cosh \left(L
   \sqrt{\frac{s}{D}}\right)}\left(1+\frac{\beta}{\sqrt{sD}}\sinh\left(\frac{L}{2}\sqrt{\frac{s}{D}}\right)\right)\right]\,.\label{eq:Sbox}
\end{align}

\begin{figure}[htbp]
  \begin{center}
    \includegraphics[width=0.6\textwidth]{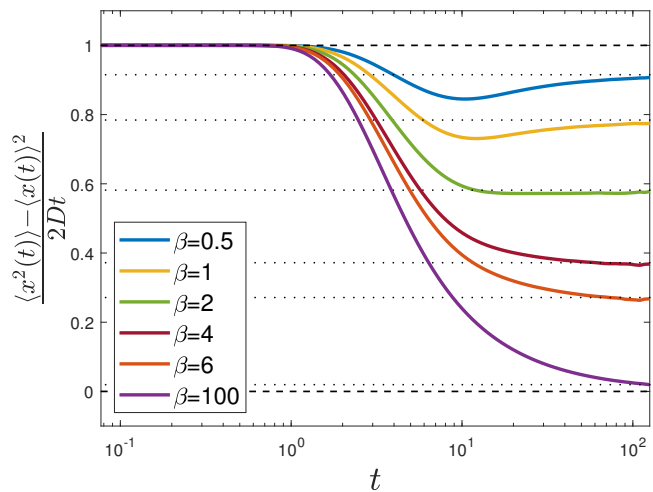}
    \caption{Time evolution of the normalised mean-square displacement per unit time for a diffusive particle starting from the origin and surviving in the presence of an infinite number of partially absorbing traps with intensity $\beta$, separated by a distance $L=10$ and evenly spaced around the initial position. Dotted lines represent the theoretically predicted effective diffusion coefficient (\ref{eq:Deffinf}) for each value of $\beta$. Colored lines are numerical data for $ 3\cdot 10^8$ Gaussian random walks evolved up to step $2\cdot 10^5$, with time spacing $\tau=0.001$, $D=1$, variance $2\tau$ and periodic trapping intervals of length $\beta\tau$.}
    \label{fig:InfTraps}
  \end{center}
\end{figure}

Let us now return to the computation of the effective diffusion coefficient in (\ref{eq:Deffint}). We will analyse the numerator and the denominator in the long time limit separately. Let us start with the numerator. Using (\ref{eq:tR}), we find that its Laplace transform is given by
\begin{align}
  \int_0^\infty dt \,e^{-st}\sum_{k=-\infty}^\infty k^2 R(k,t) = \sum_{k=-\infty}^\infty k^2 \tilde R(k,s) = \frac{\tilde f(s) \tilde S_\text{box}(s)}{[1-\tilde f(s)]^2}\,.
\end{align}
Inserting the expressions of $\tilde f(s)$ and $\tilde S_\text{box}(s)$ from (\ref{eq:fs})-(\ref{eq:Sbox}), we get
\begin{align}
  \sum_{k=-\infty}^\infty k^2 \tilde R(k,s)  =  \frac{\beta  \sqrt{D s} \tanh \left(\frac{L}{4}  \sqrt{\frac{s}{D}}\right)+2 D s}{s\beta^2 \left( \cosh \left(\frac{L}{2}  \sqrt{\frac{s}{D}}\right)+\frac{2 \sqrt{sD } }{\beta}\sinh
   \left(\frac{L}{2}  \sqrt{\frac{s}{D}}\right)\right)^2}\,.\label{eq:sumk2s}
\end{align}
In order to obtain the long time limit of $\sum_{k=-\infty}^\infty k^2 R(k,t)$, we analyse the poles of (\ref{eq:sumk2s}) in the complex $s$-plane. To do so, it is convenient to define the function 
$g(s)$ as
\begin{align}\label{eq:gs}
  g(s) \coloneqq \cosh\left(\frac{L}{2}\sqrt{\frac{s}{D}}\right)+\frac{2\sqrt{sD}}{\beta}\sinh\left(\frac{L}{2}\sqrt{\frac{s}{D}}\right)\,,
\end{align}
so that (\ref{eq:sumk2s}) reads
\begin{align}
  \sum_{k=-\infty}^\infty k^2 \tilde R(k,s)  =  \frac{\beta  \sqrt{D s} \tanh \left(\frac{L}{4}  \sqrt{\frac{s}{D}}\right)+2 D s}{s\beta^2 g(s)^2} \label{eq:sumk2sb}\,.
\end{align}
Let $-\alpha(\beta,L)<0$ be the largest negative zero of $g(s)$ such that
 \begin{align}
   g(-\alpha(\beta,L)) = 0\,.
 \end{align}
 This zero is the same as the one found in \cite{GPBD21} and is given by
 \begin{equation} 
 \alpha(\beta,L) =\frac{D}{L^2} \,\mathcal{G}\left(\mathcal{W}=\frac{\beta L}{D}\right)\,,\label{eq:alphascal}
 \end{equation}
 where $\mathcal{W}$ is the Sherwood number and the function $\mathcal{G}$ is provided in (\ref{eq:gu}).
 As $g(s)$ is linear around this point, we can write 
  \begin{align}
   g(s) \sim g'(-\alpha(\beta,L))(s+\alpha(\beta,L))\,,\qquad \mbox{as}\quad s\to -\alpha(\beta,L)\,,\label{eq:gsa}
 \end{align}
 where $g'(s)$ is the derivative of $g(s)$.
Inserting this expansion in (\ref{eq:sumk2sb}), we get
\begin{align}
  \sum_{k=-\infty}^\infty k^2 \tilde R(k,s)  \sim \frac{\beta  \sqrt{D \alpha(\beta,L)} \tan \left(\frac{L}{4}  \sqrt{\frac{\alpha(\beta,L)}{D}}\right)+2 D \alpha(\beta,L)}{\alpha(\beta,L)\beta^2 g'(-\alpha(\beta,L))^2\left(s+\alpha(\beta,L)\right)^2}\,,\qquad s\to -\alpha(\beta,L)\,.
\end{align}
Inverting this Laplace transform, we find that the numerator in the effective diffusion coefficient (\ref{eq:Deffint}) behaves, for $t\to \infty$, as
\begin{align}
\sum_{k=-\infty}^\infty k^2 R(k,t)  \sim  \frac{\left(\beta  \sqrt{D \alpha(\beta,L)} \tan \left(\frac{L}{4}  \sqrt{\frac{\alpha(\beta,L)}{D}}\right)+2 D \alpha(\beta,L)\right)t\,e^{-\alpha(\beta,L) t}}{\alpha(\beta,L)\beta^2 g'(-\alpha(\beta,L))^2} \,.\label{eq:asnum}
\end{align}
Next, we analyse the infinite sum in the denominator of (\ref{eq:Deffint}) in the long time limit. Using (\ref{eq:tR}), we obtain that the Laplace transform of the infinite sum is given by
\begin{align}
  \int_0^\infty dt e^{-st}\sum_{k=-\infty}^\infty R(k,t) = \sum_{k=-\infty}^\infty \tilde R(k,s) =  \frac{\tilde S_\text{box}(s)}{1-\tilde f(s)}\,.
\end{align}
Inserting the expressions of $\tilde f(s)$ and $\tilde S_\text{box}(s)$ from (\ref{eq:fs})-(\ref{eq:Sbox}), we get
\begin{align}
   \sum_{k=-\infty}^\infty \tilde R(k,s) = \frac{1}{s}\left(1-\frac{1}{g(s)}\right)\,,\label{eq:mS}
\end{align}
where $g(s)$ is defined in (\ref{eq:gs}).
Note that the expression (\ref{eq:mS}) is the Laplace transform of the survival probability (\ref{eq:SrelRk}) and consistently matches with the expression $(65)$ found in \cite{GPBD21} with $\theta=\pi$. Substituting the expansion (\ref{eq:gsa}) of $g(s)$ in (\ref{eq:mS}), we find that
\begin{align}
  \sum_{k=-\infty}^\infty \tilde R(k,s)\sim \frac{1}{\alpha(\beta,L) g'(-\alpha(\beta,L))(s+\alpha(\beta,L))} \,,\qquad \mbox{as}\quad s\to -\alpha(\beta,L)\,.
 \end{align}
Inverting this Laplace transform, we obtain that the series in the denominator of the effective diffusion coefficient (\ref{eq:Deffint}) behaves, for $t\to\infty$, as
 \begin{align}
  \sum_{k=-\infty}^\infty  R(k,t) \sim \frac{e^{-\alpha(\beta,L) t}}{\alpha(\beta,L) g'(-\alpha(\beta,L))}\,,\quad t\to\infty\,.\label{eq:asden}
 \end{align}
Finally, inserting the asymptotic behavior of the numerator (\ref{eq:asnum}) and the one of the series in the denominator (\ref{eq:asden}) into (\ref{eq:Deffint}), we can conclude that the effective diffusion coefficient is given by
\begin{align}
 D_\text{eff} =\lim_{t\to\infty}\,\frac{L^2}{2t}\,  \frac{\langle k^2(t)\rangle}{S(t)} = L^2\frac{\left(\beta  \sqrt{D \alpha(\beta,L)} \tan \left(\frac{L}{4}  \sqrt{\frac{\alpha(\beta,L)}{D}}\right)+2 D \alpha(\beta,L)\right)}{2\beta^2 g'(-\alpha(\beta,L))} \,.
\end{align}
Using the explicit expression for the derivative of $g(s)$ in (\ref{eq:gs}) and further simplifying knowing that $g(-\alpha(\beta,L))=0$, we finally obtain the result (\ref{eq:Deff}) announced in the introduction. The time evolution of the normalised mean-square displacement per unit time is shown in figure~\ref{fig:InfTraps}. In the long time limit, it converges to the predicted effective diffusion coefficient in (\ref{eq:Deffinf}). At short times, the particle undergoes normal diffusion, since it is not yet aware of the presence of the point absorbers. Then, at intermediate times, such that the first traps are encountered but have not yet been uniformly sampled, anomalous diffusion occurs. At long times, normal diffusion is recovered, although with a different diffusion coefficient, witnessing the creation of a new equilibrium (averaged mobility) between particles and traps. The anomalous diffusion is therefore apparent and turns out to be of transitional type. It corresponds to the inflection point of the continuous interpolation between fast and slow normal diffusion \cite{Berthier2005,biological-transient-subdiffusion,biological-transient-subdiffusion-bis}. 
It should be stressed that this transitional anomalous regime has to be distinguished from the true transient anomalous diffusion exemplified in the literature, for instance,
by a random walk on obstructed lattices \cite{obstacles-transient-subdiffusion,obstacles-transient-subdiffusion2}. 

\subsection{Effective Langevin equation}
In this section, we provide an effective Langevin equation to generate surviving trajectories in an efficient manner. The construction presented in this section is known as a Doob transform in the probability literature \cite{Doob,Pitman} and has generated recent interest in the physics community due to its various applications \cite{BCDG2002,GKP2006,GKLT2011,KGGW2018,Gar2018,Rose21,Rose21area,CLV21,CT2013,MajumdarEff15,Orland,Grela2021,DebruyneAR21,Mazzolo17a,Mazzolo17b,Monthus21,Baldassarri21,DebruyneRW21,DebruyneRTP21,Brunet20}.
The goal is to generate trajectories of surviving particles up to a final time $t_f$. To do so, we write the \emph{constrained} propagator $p_c(x,t\,|\,t_f)$ for the surviving particle, i.e. the probability distribution for the surviving particle to be located at position $x$ at an intermediate time $t$. Due to the Markov nature of the process, it reads
\begin{align}
  p_c(x,t\,|\,t_f) = \frac{p(x,t)S(t_f-t|x)}{S(t_f|x=0)}\,,\label{eq:pc}
\end{align}
where $p(x,t)$ accounts for all the trajectories going for the origin to the point $x$ at time $t$, $S(t_f-t|x)$ accounts for all the trajectories that survive from $x$ during the remaining time $t_f-t$, and the denominator is the normalisation coefficient that counts all the trajectories that survived from the origin. The propagator $p(x,t)$ satisfies the Fokker-Planck equation (\ref{eq:fp}). The survival probability $S(t|x)$ satisfies the backward Fokker-Planck equation 
\begin{align}
  \partial_t S(t|x) = D \partial_{xx} S(t|x)- \beta \sum_{m=-\infty}^{\infty} \delta\left(x-\frac{L}{2}-m\,L\right)S(t|x)\,,\label{eq:Sp}
\end{align}
with the initial condition $S(0|x)=1$. By taking a time derivative of the constrained propagator in (\ref{eq:pc}), and using the two Fokker-Planck equations (\ref{eq:fp}) and (\ref{eq:Sp}), one can show that the constrained propagator satisfies the effective Fokker-Planck equation 
\begin{align}
  \partial_t p_c(x,t|t_f) = D\partial_{xx} p_c(x,t|t_f) - 2D \partial_x\left[\partial_x\left(\ln(S(t_f-t|x))\right) p_c(x,t|t_f)\right]\,.\label{eq:fpeff}
\end{align}
Note that, compared to (\ref{eq:Sp}) and (\ref{eq:fp}), the absorbing terms are no more present, and an additional force term appeared.
One observes that the effective Fokker-Planck equation (\ref{eq:fpeff}) can be obtained from an effective Langevin equation for the trajectory $x_s(t)$ of surviving particles given by
\begin{align}
  \dot x_s(t) = \sqrt{2D}\,\eta(t) + 2D \partial_{x_s} \ln[S(t_f-t|x_s)]\,,\label{eq:eomeff}
\end{align}
where $\eta(t)$ is a Gaussian white noise with zero mean $\langle \eta(t)\rangle=0$ and delta-correlations $\langle \eta(t)\eta(t')\rangle=\delta(t-t')$. The subscript $s$ in $x_s(t)$ refers to ``surviving'' trajectories. The survival probability $S(t|x)$ is the solution of the differential equation (\ref{eq:Sp}) and was computed in \cite{GPBD21} (see equation (65) therein). In our notation, its Laplace transform reads
\begin{align}
\tilde S(s|x) =  \frac{1}{s}\left(1-\frac{\cosh\left(x\sqrt{\frac{s}{D}}\right)}{g(s)}\right)\,,\qquad -\frac{L}{2}\leq x\leq \frac{L}{2}\,,\label{eq:Stx}
\end{align}
and is $L$-periodic $\tilde S(s|x)=\tilde S(s|x+L)$, with $g(s)$ given in (\ref{eq:gs}).
This Laplace transform is difficult to invert for arbitrary $t$. However, we can study its long time limit. Performing a similar computation as in the previous section, we find that the long time limit is dominated by the behavior of $\tilde S(s|x)$ close to $s=-\alpha(\beta,L)$, which is the largest negative zero of $g(s)$. Close to this point, the Laplace transform behaves as
\begin{align}
  \tilde S(s|x) \sim \frac{1}{\alpha(\beta,L)}\frac{\cos\left(x\sqrt{\frac{\alpha(\beta,L)}{D}}\right)}{g'(-\alpha(\beta,L))(s+\alpha(\beta,L))}\,,\qquad -\frac{L}{2}\leq x\leq \frac{L}{2}\,.\label{eq:Stix}
\end{align}
Inverting this Laplace transform and taking the derivative of its logarithm, we find that
\begin{align}
  \partial_x \ln(S(t|x)) \sim \sqrt{\frac{\alpha(\beta,L)}{D}} \tan\left(x\sqrt{\frac{\alpha(\beta,L)}{D}}\right)\,,\qquad -\frac{L}{2}\leq x\leq \frac{L}{2}\,,\qquad t\to \infty.\label{eq:effF}
\end{align}
Using the expression (\ref{eq:effF}), we recover the effective Langevin equation (\ref{eq:effLf}) with the effective potential (\ref{eq:Ueff}). In (\ref{eq:Ueff}), we used that $\alpha(\beta,L)=D/L^2 \mathcal{G}(\mathcal{W})$ where $\mathcal{G}(\mathcal{W})$ is given in (\ref{eq:gu}). Note that in the limit $\beta \to \infty$, using the asymptotic expansion $\mathcal{G}(\mathcal{W})\sim \pi^2$ for $\mathcal{W}\to \infty$ \cite{GPBD21}, the effective potential becomes
\begin{align}
  U_\text{eff}(x) \sim -2D \ln\left[\cos\left(\frac{x \pi}{L}\right)\right]\,,\qquad -\frac{L}{2}\leq x\leq \frac{L}{2}\,,\qquad \beta \to \infty\,,\label{eq:Ueffl}
\end{align} 
which diverges at the locations of the traps and can be used to generate surviving trajectories in the presence of fully absorbing traps.

\begin{figure}[htbp]
  \begin{center}
    \includegraphics[width=0.6\textwidth]{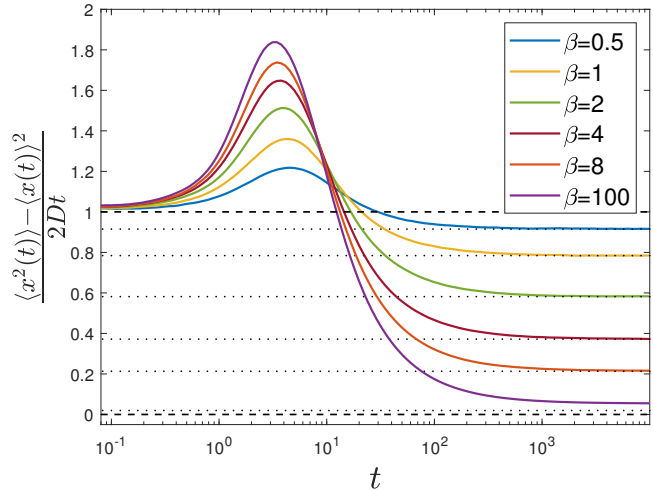}
    \caption{Time evolution of the normalised mean-square displacement per unit time for a diffusive particle subject to a potential of the form \eqref{eq:Ueff}, with parameters $D=1$ and $L=10$. Dotted lines represent the theoretically predicted effective diffusion coefficient (\ref{eq:Deffinf}) for each value of $\beta$. Colored lines are numerical results for $ 10^6$ diffusive particles evolved up to time $10^4$, where the temporal discretization of \eqref{eq:effLf} has been performed with a time step $\tau=0.001$.}
    \label{fig:InfTrapsEff}
  \end{center}
\end{figure}

One can then discretise the effective Langevin equation (\ref{eq:effLf}) over small time increments and generate trajectories of particles that survive for $t_f\to\infty$. In figure \ref{fig:InfTrapsEff}, we illustrate the evolution of the rescaled mean-square displacement of such particles for various trap intensities $\beta$. For $t\to\infty$, it converges to the predicted value in (\ref{eq:Deffinf}). Note that the full time dependence is different from the one obtained in figure \ref{fig:InfTraps}. This is due to the fact that in figure \ref{fig:InfTraps}, the conditioning is done on the particles that are still alive at time $t$, whereas in figure \ref{fig:InfTrapsEff}, the conditioning is performed on the particles that are still alive at a future time $t_f\to \infty$.

\section{Generalisation to other trapping environments}\label{Sec: various}
In the previous section, we showed that conditioning a particle to survive in a symmetric trapping environment can induce an effective diffusion coefficient that is different from the original one. In this section, we discuss the case of an asymmetric trapping environment and highlight several features, in particular that it induces a drift on the surviving particle. We discuss in details the cases of a single trap and of infinite number of traps on the positive semi-axis.
\subsection{Single trap }
\begin{figure}[htbp]
  \begin{center}
    \includegraphics[width=0.48\textwidth]{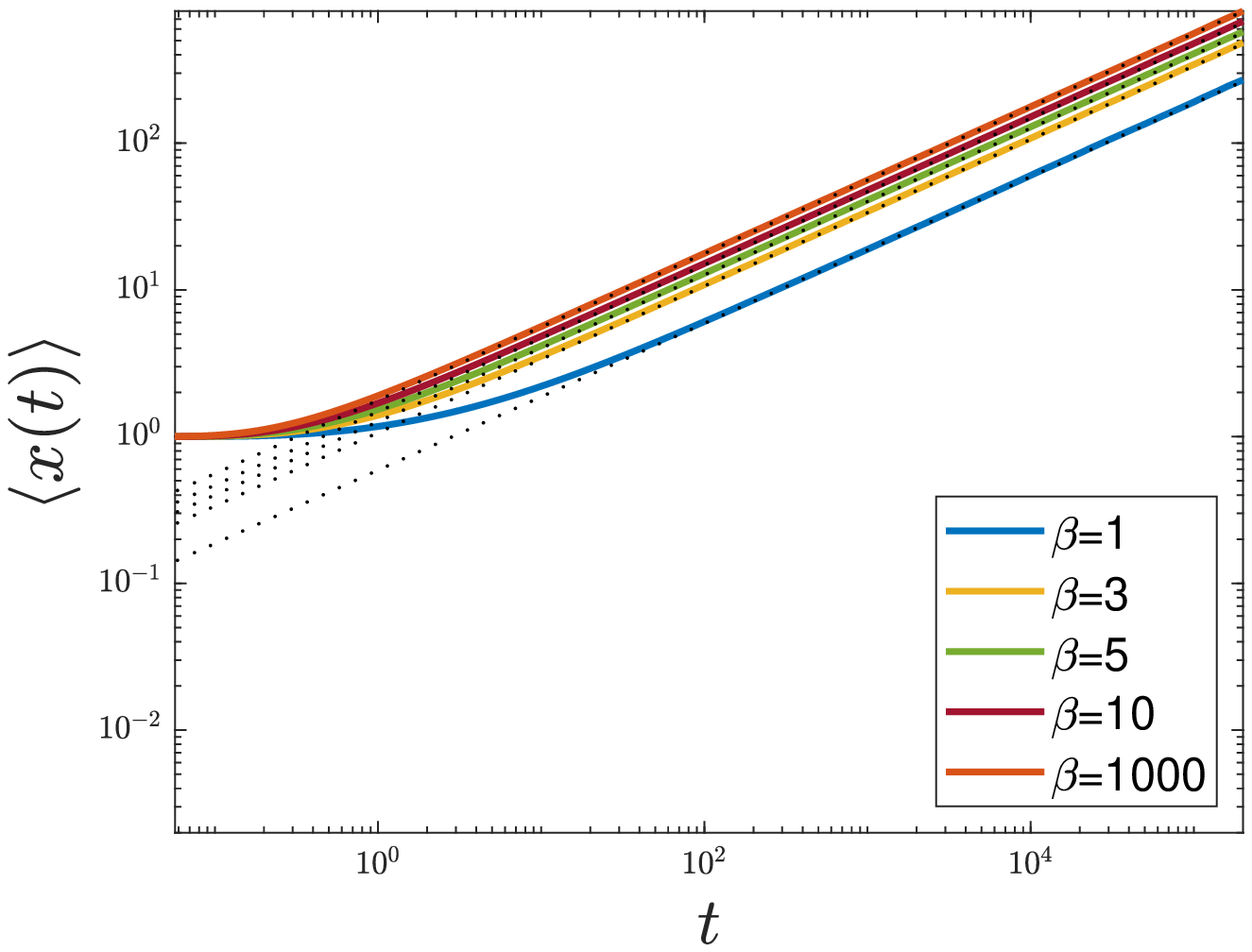}
\includegraphics[width=0.48\textwidth]{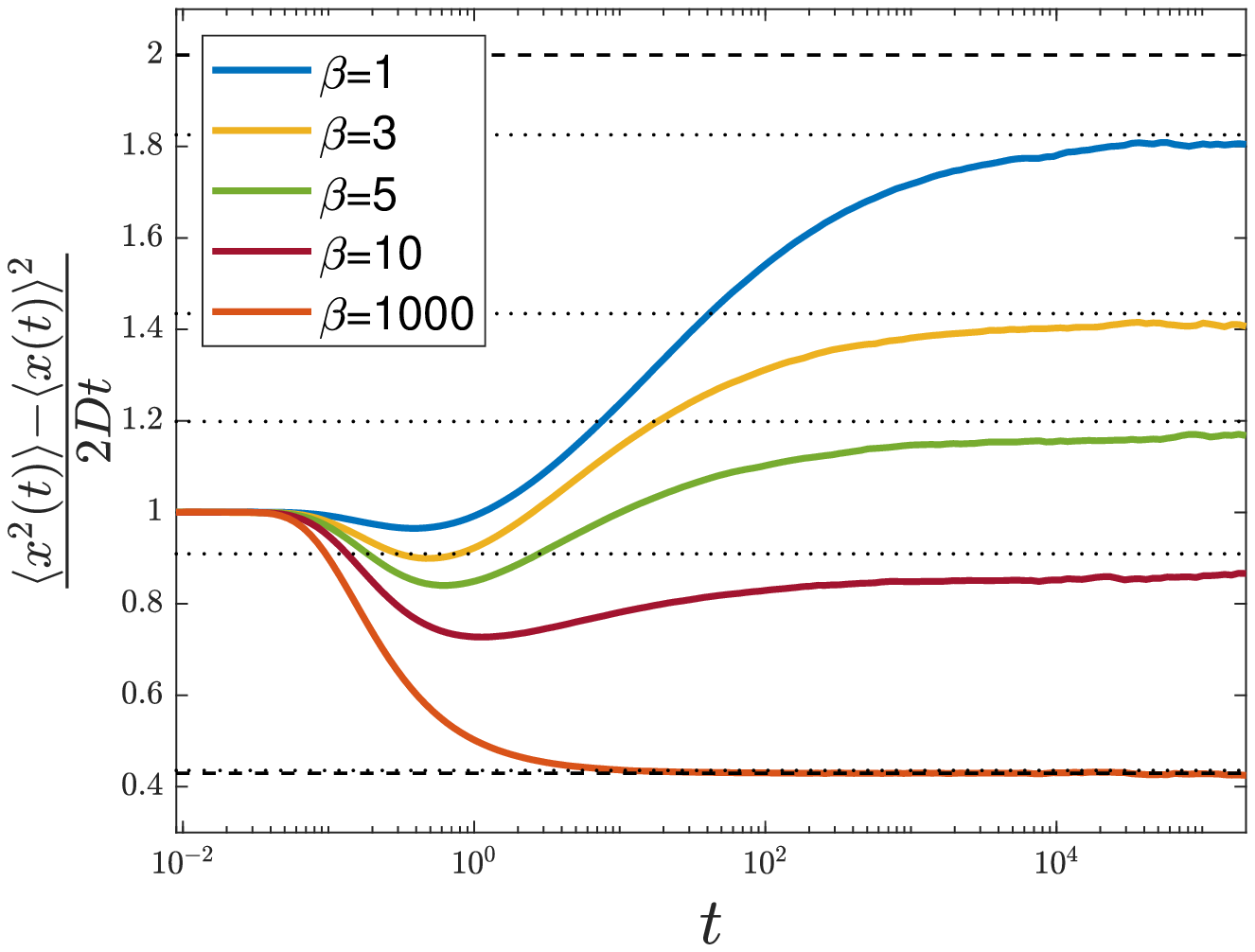}
    \caption{Time evolution of the mean drift (left panel) and of the normalised mean-square displacement per unit time (right panel) for a diffusive particle, starting from $x_0=1$, which survives in the presence of a single point absorber with intensity $\beta$ located at the origin. Dotted lines in the first panel depict the theoretical current in \eqref{eq:asAsym} for each value of $\beta$, whereas dotted lines in the second panel represent the theoretically predicted effective diffusion coefficients in (\ref{eq: ratio}). Colored lines are numerical data for $  3\cdot 10^7$ Gaussian random walks evolved up to step $2\cdot 10^8$, with time increment $\tau=0.001$, $D=1$, variance $2\tau$ and a trapping interval of length $\beta\tau$ at the origin.}
    \label{fig:1TrapAsym}
  \end{center}
\end{figure}

In this section, we discuss the case of an asymmetric trapping environment composed of a single trap. The Fokker-Planck equation is given in (\ref{eq:fp1}) with the initial condition $p(x,t=0)=\delta(x-x_0)$ with $x_0>0$. Let us first show that the environment induces a drift on the surviving particle. To do so, we compute the conditional first moment in (\ref{eq:sm}). Using the expression of the first moment in (\ref{eq:fmom}) and the fact that the survival probability in (\ref{eq:Sb}) for large $t$ behaves as
\begin{align}
  S(t) \sim \left(\frac{2\sqrt{D}}{\beta \sqrt{\pi}} + \frac{x_0}{\sqrt{\pi D}}\right)\frac{1}{\sqrt{t}}\,,\qquad t \to \infty\,,\label{eq:Sa}
\end{align}
we find that the conditional first moment behaves as
\begin{align}
 \langle  x(t)\rangle \sim \left\{\begin{array}{ll}
    x_0 &\qquad t\to 0\,,\\
    \frac{x_0\beta \sqrt{\pi D t}}{2D+ \beta x_0} & \qquad t\to \infty\,.
  \end{array}\right. \label{eq:asAsym}
\end{align}
Therefore, the mean position of the surviving particle grows as $\sqrt{t}$ for large $t$ with an amplitude that depends non-trivially on the initial position and the trap intensity $\beta$. This result is in good agreement with simulations, as can be seen in figure \ref{fig:1TrapAsym}.

Let us now compute the mean-square displacement in (\ref{eq:DeffGen}). Upon using the expression of the second moment in (\ref{eq:x2}) and the expansion of the survival probability in (\ref{eq:Sa}), we find that the  conditional mean-square displacement behaves as
\begin{align}
  \frac{ \langle  x(t)^2\rangle - \langle  x(t)\rangle^2 }{2t} \sim \left\{\begin{array}{ll}
    D &\qquad t\to 0\,,\\
    \left(2-\frac{\pi  \beta ^2 x_0^2}{2(2 D+\beta x_0)^2}\right)D  & \qquad t\to \infty\,.
  \end{array}\right. \label{eq: ratio}
\end{align}
 In particular, note that in contrast to the symmetric configurations with a finite number of traps, the effective diffusion is not necessarily enhanced. This result is in good agreement with numerical simulations (see figure \ref{fig:1TrapAsym}).

In the small $\beta$ limit, it is also possible to analytically characterise the presence of a cusp in the probability density function in correspondence with the position of the point absorber.
In the limit $\beta\to 0$, the integration constants in (\ref{eq:ABCE}) at first order are given by
\begin{align}\label{eq:ABCEa}
   A(s) &\sim \frac{e^{-\sqrt{\frac{s}{D}} x_0}}{2\sqrt{s D}}\left(1 - \frac{\beta}{2 \sqrt{sD}}\right)+O(\beta^2)\,,\quad  B(s) \sim \frac{e^{-\sqrt{\frac{s}{D}} x_0}}{2\sqrt{s D}}+O(\beta^2)\,,\nonumber\\
   \quad C(s) &\sim - \beta \frac{e^{-\sqrt{\frac{s}{D}} x_0}}{4sD}+O(\beta^2)\,,    \quad  E(s) \sim \frac{e^{\sqrt{\frac{s}{D}} x_0}}{2\sqrt{s D}} - \beta \frac{e^{-\sqrt{\frac{s}{D}} x_0}}{4sD}+O(\beta^2)\,.
\end{align}
Plugging these asymptotic behaviors in (\ref{eq:gens}) and Laplace inverting, we get
\begin{align}
  p(x,t) \sim \frac{1}{\sqrt{4\pi D t}} e^{-\frac{(x-x_0)^2}{4Dt}} - \frac{\beta}{4D} \text{erfc}\left(\frac{x_0+|x|}{\sqrt{4Dt}}\right)\,,\qquad \beta\to 0\,.\label{eq:pa}
\end{align}
 The first term is the usual Gaussian profile of free diffusion on the real line, whereas the second term accounts for the correction associated with the trap. The absolute value in the second term in (\ref{eq:pa}) is clearly responsible for the cusp behavior. 

Note that since $x_0$ does not coincide with the trap, the initial pure diffusion stays during a time of order $t\lesssim x_0^2/4D$, which is consistent with the fact that the crossover time must be independent of the value of $\beta$ since it depends mostly on the distance from the point absorber. If $x_0=0$, as we have seen in section~\ref{subsec:single}, this leading term vanishes and the secondary effect, characteristic of the trap intensity, becomes evident, as shown in figure~\ref{fig:1TrapOrigin}.

\subsection{Infinite number of traps on the positive semi-axis}
In analogy with the single trap configuration, it is also possible to induce a permanent current by placing an infinite number of periodically distributed point absorbers on one side only, or similarly by having different trap intensities to the left and to the right of the initial position of the particle. Here we will focus specifically on the former example, where a similar argument to that presented in section \ref{sec:arbFinite} can provide a direct estimate. 

\begin{figure}[htbp]
  \begin{center}
    \includegraphics[width=0.48\textwidth]{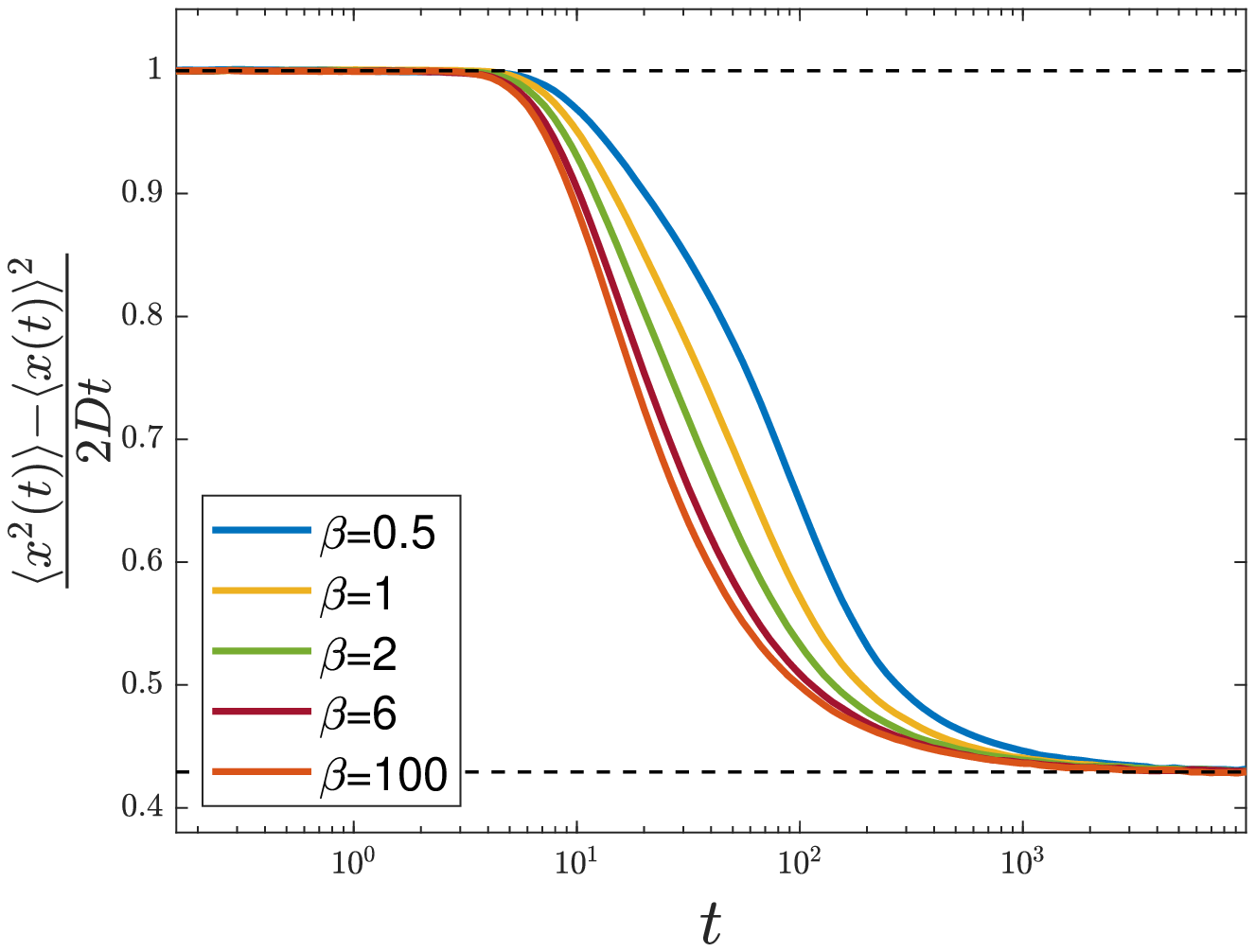}
    \includegraphics[width=0.48\textwidth]{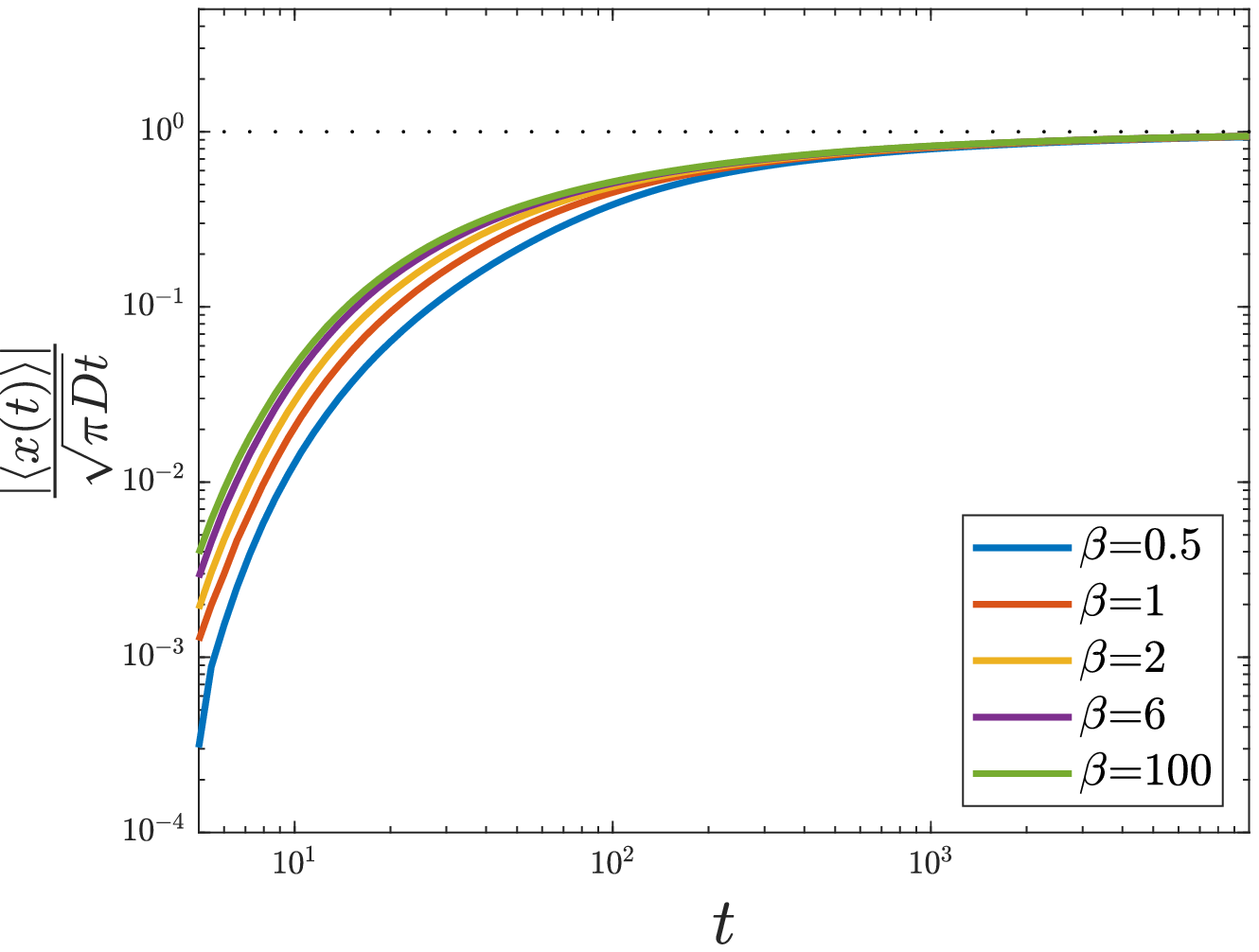}
    \caption{Time evolution of the normalised mean-square displacement per unit time (left panel) and of the mean drift (right panel) for a diffusive particle, starting from $x_0=0$, which survives in the presence of infinitely many unilateral absorbing traps with intensity $\beta$ and periodically separated by a distance $L=10$ to the right of the origin. The lower dashed line in the first panel represent the theoretical effective diffusion coefficient in \eqref{eq:unilDiff}, whereas in the second panel we show that the ratio between the numerical induced current and the leading order term of the prediction in \eqref{eq:unilDrift} converges to unity. Colored lines are data from simulations of $  10^7$ Gaussian random walks evolved up to step $10^7$, with time spacing $\tau=0.001$, $D=1$, variance $2\tau$ and periodic trapping intervals of length $\beta\tau$.}
    \label{fig:unil}
  \end{center}
\end{figure}

Let us consider a Brownian motion starting from the origin in the presence of an infinite number of point absorbers located at $jL$ with $j\in\mathbb{N}\,$. In the long time limit, we expect that the infinite unilateral array of traps effectively behaves as a fully absorbing barrier, which is confirmed in figure \ref{fig:unil}. More explicitly, in order to determine the asymptotic drift and the effective diffusion coefficient in the presence of infinite unilateral traps with respect to the origin, it is sufficient to perform the limit $\beta\to\infty$ in \eqref{eq:asAsym} and \eqref{eq: ratio} respectively, and so assuming that the particle starts from $x_0\leq L$ we can write
\begin{eqnarray}
&\langle x(t)\rangle\sim -\sqrt{\pi D t}\,,\qquad &\mbox{as}\quad t\to \infty\,,\label{eq:unilDrift}\\
&D_\text{eff} \sim \left(2-\frac\pi 2 \right)D\,,\qquad& \mbox{as}\quad t\to \infty\,,\label{eq:unilDiff}
\end{eqnarray}
where in the first line one has to take into account a reflection $x\to -x$ with respect to the original setup of the previous section as the semi-infinite array of traps is located on the positive axis.

\section{Conclusion}

In this paper, we investigated the transport properties of a one-dimensional Brownian particle conditioned to survive in the presence of periodically distributed point absorbers with a fixed trap intensity $\beta\,$. We showed that, in the long time limit, the system attains a normal diffusive regime. More specifically, as long as the trapping region is bounded and symmetric around the starting position, the effective diffusion coefficient is universally equal to twice the original one. In the presence of an infinite array of absorbing traps, instead, the asymptotic mean-square displacement per unit time is a non-trivial function of the Sherwood number, which encloses the interlacing between the distance among the point absorbers, the original diffusion coefficient and the trapping strength. Furthermore, we provided a rejection-free algorithm to generate surviving trajectories. As a final step, we extended the analysis to asymmetric environments.

In this work we provided rough estimates on the order of magnitude of the crossover times, nevertheless it could be worth determining the exact separation of time scales, by identifying more precisely the time windows for short and long time regimes. 
Moreover, beyond the asymptotic regime, it would be interesting to characterize the transient anomalous diffusion at intermediate times, and in particular the existence of extrema in the evolution of the mean-square displacement. 

\section*{Acknowledgements}
GP warmly thanks Roberto Artuso for the helpful initial discussion. GP acknowledges support from PRIN Research Project No. 2017S35EHN ``Regular and stochastic behavior in dynamical systems" of the Italian Ministry of Education, University and Research (MIUR). BD warmly thanks Gr\'egory Schehr and Satya N. Majumdar for fruitful comments and feedback. BD gratefully acknowledges the financial support of the Luxembourg National
Research Fund (FNR) (App. ID 14548297).

\appendix
\section{Two Traps}\label{sec:two}
\begin{figure}[htbp]
  \begin{center}
    \includegraphics[width=0.6\textwidth]{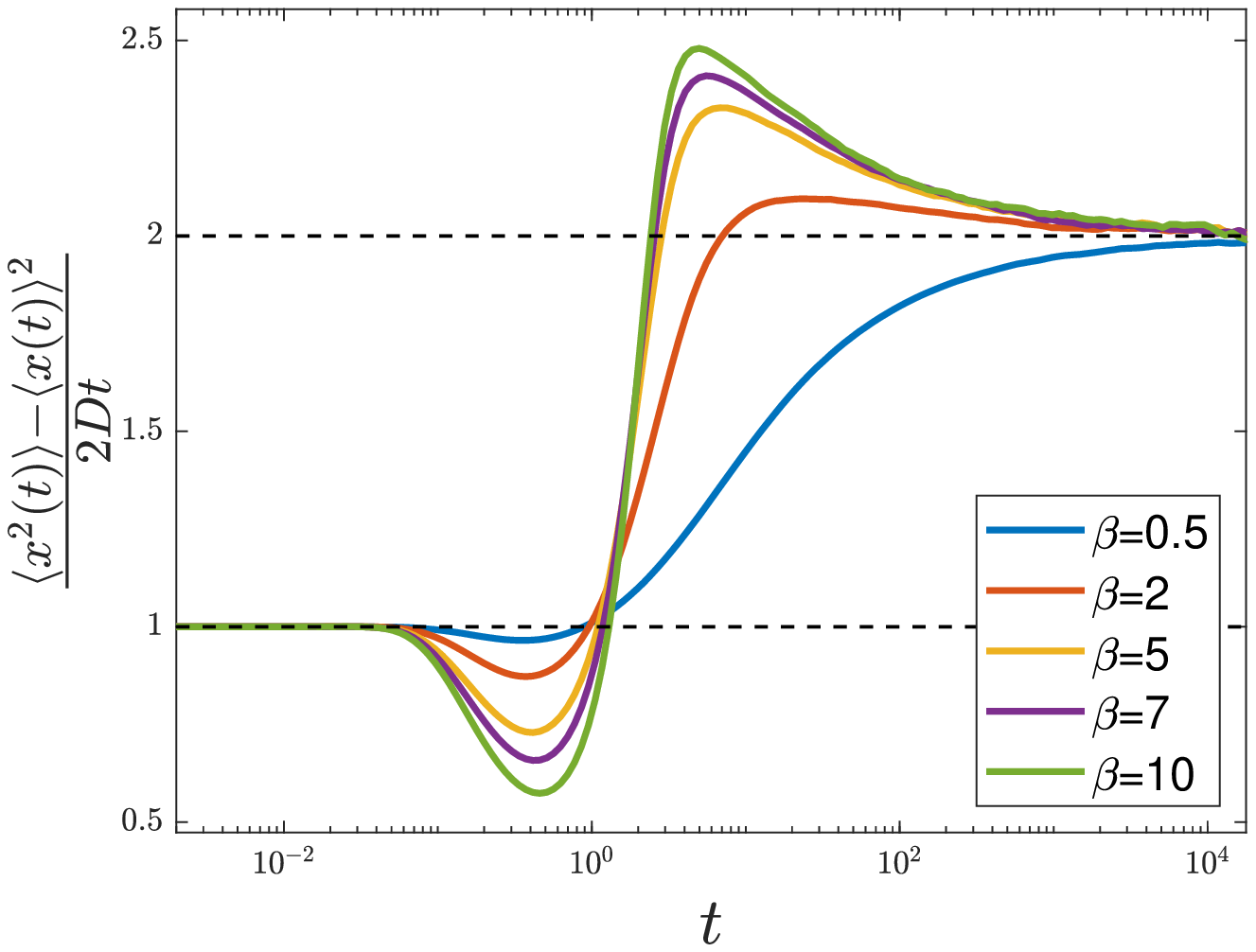}
    \caption{Time evolution of the normalised mean-square displacement per unit time for a diffusive particle with initial condition $x_0=0$ that survives in the presence of two partially absorbing traps with intensity $\beta$ located at $x=\pm1$. Dashed lines represent the theoretically predicted effective diffusion coefficient \eqref{eq:as2} at short and long times. Numerical results are obtained by simulating $ 5\cdot 10^7$ Gaussian random walks up to step $5\cdot 10^7$, with time increment $\tau=0.001$, $D=1$, variance $2\tau$ and trapping intervals of length $\beta\tau$.}
    \label{fig:2traps}
  \end{center}
\end{figure}

We consider a one-dimensional Brownian motion in the presence of two partially absorbing traps located at $\pm x_1$. The probability distribution $p(x,t)$ of the position $x$ of the Brownian motion at time $t$ follows the forward Fokker-Planck equation
\begin{align}
  \partial_t p(x,t) = D \partial_{xx} p(x,t) -\beta \delta(x-x_1) p(x,t)-\beta \delta(x+x_1) p(x,t)\,,\label{eq:fp22}
\end{align}
where $D$ is the diffusion constant and $\beta$ is the trap intensity. We assume that the particle is initially at $x_0=0$ with $p(x,0)=\delta(x)$ and that $x_1>0$. In Laplace domain (\ref{eq:fp22}) writes
\begin{align}
  s \tilde p(x,s) -\delta(x) = D \partial_{xx}\tilde p(x,s) -\beta \delta(x-x_1)\tilde  p(x,s)-\beta \delta(x+x_1)\tilde  p(x,s)\,,\label{eq:fp2s}
\end{align}
where $\tilde  p(x,s)  = \int_0^\infty dt e^{-st}p(x,t)$. The equation (\ref{eq:fp2s}) can be solved on four separate intervals. The general solution is of the form \begin{align}
  \tilde p(x,s) = \left\{\begin{array}{ll}
    A(s) e^{x \sqrt{\frac{s}{D}}}\,,\qquad & x< -x_1\,,\\
    B(s) e^{x \sqrt{\frac{s}{D}}} + C(s)  e^{-x \sqrt{\frac{s}{D}}}\,, \qquad& -x_1<x<0\,,\\
     E(s) e^{x \sqrt{\frac{s}{D}}} + F(s)  e^{-x \sqrt{\frac{s}{D}}}\,, \qquad& 0<x<x_1\,,\\
    G(s) e^{-x \sqrt{\frac{s}{D}}}\,, \qquad& x_1<x\,.
  \end{array}\right.\label{eq:gens2}
\end{align}
By imposing the continuity of the solution and matching the derivative by integrating (\ref{eq:fp2s}) in a neighborhood of the points located at the interfaces between the four regions, we find that the integration constants are uniquely determined by
\begin{subequations}
\begin{align}
  A(s) &= G(s) =\frac{1}{\beta +\beta  e^{-2 x_1 \sqrt{\frac{s}{D}}}+2
   \sqrt{D s}}\,,\\
   B(s) &= F(s) =\frac{\left(\beta ^2-4 D s\right) }{2 D \left[\left(\beta ^2
   \sqrt{\frac{s}{D}}-4 \sqrt{D s^3}\right)+\beta  \left(\beta 
   \sqrt{\frac{s}{D}}-2 s\right)e^{-2 x_1
   \sqrt{\frac{s}{D}}}\right]}\,,\\
   C(s) &= E(s) =-\frac{\beta }{2 D \left[\beta 
   \sqrt{\frac{s}{D}}+\left(\beta  \sqrt{\frac{s}{D}}+2
   s\right) e^{2 x_1 \sqrt{\frac{s}{D}}}\right]}\,.
\end{align}
\label{eq:ABCDEFG2}
\end{subequations}
The Laplace transform of the survival probability is therefore given by
\begin{align}
  \tilde S(s) =  \int_{-\infty}^{\infty} dx \,\tilde p(x,s) = \frac{1}{s}-\frac{2 \beta  e^{x_1 \sqrt{\frac{s}{D}}}}{s \left[\beta +\left(\beta +2 D \sqrt{\frac{s}{D}}\right) e^{2 x_1
   \sqrt{\frac{s}{D}}}\right]}\,,\label{eq:Ss2}
\end{align}
and we can also compute the Laplace transform of the second moment 
\begin{align}
 \int_{-\infty}^\infty dx x^2 \tilde p(x,s) = \frac{2D}{s}-\frac{2 \beta  e^{x_1 \sqrt{\frac{s}{D}}} \left(2 D+s x_1^2\right) \left[\left(2 D \sqrt{\frac{s}{D}}-\beta \right) e^{2 x_1
   \sqrt{\frac{s}{D}}}-\beta \right]}{s^2 \left[4 D s e^{4 x_1 \sqrt{\frac{s}{D}}}-\beta ^2 \left(e^{2 x_1
   \sqrt{\frac{s}{D}}}+1\right)^2\right]}\,.\label{eq:x2s2}
\end{align}
By considering the ratio of these two expressions \eqref{eq:x2s2} and \eqref{eq:Ss2} and taking a small $s$ limit, we finally find the asymptotic behaviour of the effective diffusion coefficient in the long time limit
\begin{align}
  \frac{ \langle  x(t)^2\rangle - \langle  x(t)\rangle^2 }{2t}  \sim \left\{\begin{array}{ll}
    D & \qquad t\to 0\,,\\
   2D &\qquad  t\to \infty\,,
  \end{array}\right. \label{eq:as2}
\end{align}
which is in agreement with figure \ref{fig:2traps}. Notice that this is the same result as in (\ref{eq: ratio}) for $x_0=0$. 

\section{First-exit probability and survival probability in a box with intermediate point absorbers}\label{app:InfBox}
As outlined in the main text, we need to compute the first-exit probability and the survival probability in the interval $[-L,L]$ with partially absorbing traps located at $\pm L/2$ and initial position $x_0=0$. By symmetry, it is equivalent to consider the first-passage event to the absorbing boundary $x=L$, starting from the reflecting boundary $x=0$, of a Brownian motion on the finite interval $[0,L]$ in the presence of a point absorber with intensity $\beta$ at $ x=L/2$.
First of all, we have to solve the forward Fokker-Planck equation
\begin{equation}\label{eq:FPbox}
\partial_tp(x,t)=D\partial_{xx}p(x,t)-\beta\delta(x-L/2)p(x,t)\,, \qquad \partial_xp(x,t)|_{x=0}=0\,,\quad p(x,t)|_{x=L}=0\,, 
\end{equation}
 for the unconditional propagator, namely the probability distribution $p(x,t)$ of the position $x$ of the Brownian motion at time $t$. In Laplace domain, assuming that the particle is initially located at $x_0\in]0,L/2[$ (we will then consider the limit $x_0\to 0$ for our purposes), namely $p(x,0)=\delta(x-x_0)$, the forward equation becomes
\begin{align}\label{eq: propLap}
  s \tilde p(x,s) -\delta(x-x_0) = D \partial_{xx}\tilde p(x,s) -\beta \delta(x-L/2)\tilde  p(x,s)\,.
\end{align}
Hence we have to look for a solution, on three separate intervals, of the form
\begin{align}\label{eq:solBox}
  \tilde p(x,s) = \left\{\begin{array}{ll}
    A(s) \cosh(x \sqrt{\frac{s}{D}})\,, \qquad& 0<x< x_0\,,\\
\vspace*{0.1cm}
    B(s) e^{x \sqrt{\frac{s}{D}}} + C(s)  e^{-x \sqrt{\frac{s}{D}}}\,, \qquad& x_0<x<\tfrac L 2\,,\\
\vspace*{0.2cm}
    E(s) \sinh((L-x) \sqrt{\frac{s}{D}})\,, \qquad& \tfrac L 2<x<L\,,\\
  \end{array}\right.
\end{align}
where in the first (respectively third) line we considered the reflecting (respectively absorbing) boundary condition introduced in \eqref{eq:FPbox}.
In order to determine the integration constants $A(s)$, $B(s)$, $C(s)$ and $E(s)$, we have to impose also the interface conditions between the three regions, that is\\
\vspace*{0.2cm}
{\it Continuity of the solution at $x=x_0$:} 
\begin{equation}
 A\cosh(x_0\sqrt{s/D})=Be^{x_0\sqrt{s/D}}+Ce^{-x_0\sqrt{s/D}}\,;
\end{equation}
\vspace*{0.2cm}
{\it Discontinuity of the derivative at $x=x_0$:} by integrating \eqref{eq: propLap} in a neighborhood of $x_0$, we immediately get $\partial_x\tilde p(x,s)|_{x_0^+}-\partial_x\tilde p(x,s)|_{x_0^-}=-1/D$, and substituting the solution \eqref{eq:solBox} we have
\begin{equation}
Be^{x_0\sqrt{s/D}}-Ce^{-x_0\sqrt{s/D}}-A\sinh(x_0\sqrt{s/D})=-1/\sqrt{sD}\,;
\end{equation}
\vspace*{0.2cm}
{\it Continuity of the solution at $x=L/2$:} 
\begin{equation}
 Be^{L/2\sqrt{s/D}}+Ce^{-L/2\sqrt{s/D}}=E\sinh(L/2\sqrt{s/D})\,;
\end{equation}
\vspace*{0.2cm}
{\it Discontinuity of the derivative at $x=L/2$:} the integration of \eqref{eq: propLap} around $L/2$ provides $\partial_x\tilde p(x,s)|_{\frac{L^+}2}-\partial_x\tilde p(x,s)|_{\frac{L^-}2}=\frac \beta D\tilde p(x,s)|_{\frac L2}$, and explicitly gives
\begin{equation}
-E\cosh(L/2\sqrt{s/D})-Be^{L/2\sqrt{s/D}}+Ce^{-L/2\sqrt{s/D}}=\frac{\beta}{\sqrt{sD}}E\sinh(L/2\sqrt{s/D})\,.
\end{equation}
Therefore the solution is uniquely identified by
\begin{subequations}
\begin{align}
  B(s) &=\frac{A(s)}2 -\frac 1{2\sqrt{sD}}e^{-x_0\sqrt{\frac s D }}\,,\\
  C(s) &=\frac{A(s)}2 +\frac 1{2\sqrt{sD}}e^{x_0\sqrt{\frac s D }}\,,\\
  A(s) &= E(s) \tanh\left(\frac L 2 \sqrt{\frac s D}\right)+\frac 1 {\sqrt{sD}}\frac{\sinh\left(\left(\frac L2-x_0\right)\sqrt{\frac s D}\right)}{\cosh\left(\frac L2\sqrt{\frac s D}\right)}\,,\\
  E(s)&=\frac 1 {\sqrt{sD}}\frac{\cosh\left( x_0 \sqrt{\frac s D}\right)}{\cosh \left(\frac L 2 \sqrt{\frac s D}\right)}\left[ \frac \beta{\sqrt{sD}}\sinh \left(\frac L 2 \sqrt{\frac s D}\right)+\cosh\left(\frac L 2 \sqrt{\frac s D}\right)+\frac{\sinh^2\left(\frac L 2 \sqrt{\frac s D}\right)}{\cosh\left(\frac L 2 \sqrt{\frac s D}\right)}\right]^{-1}\,.\label{eq:E}
\end{align}
\label{eq:ABCDEFG}
\end{subequations}
At this point, we can thus compute the Laplace transform of the first-exit probability 
\begin{eqnarray}
  \tilde f(s) &=&-D\partial_x \tilde p(x,s)\rvert_{x=L,x_0=0} = \sqrt{sD} E(s)\rvert_{x_0=0}\,.
  \end{eqnarray}
  Inserting the expression (\ref{eq:E}), we recover the expression (\ref{eq:fs}) given in the main text.
The Laplace transform of the survival probability in the box $[-L,L]$ is given by the complementary probability that the particle did not exit the box and did not get absorbed in the trap, which reads
\begin{align}
  \tilde S_\text{box}(s) &= \frac{1}{s}\left(1-\tilde f(s)-D\left[\partial_x\tilde p(x,s)|_{\frac{L^+}2}-\partial_x\tilde p(x,s)|_{\frac{L^-}2}\right]\right)\nonumber\\
  &= \frac{1}{s}\left(1-\tilde f(s)-\beta \tilde p\left(x,s\right)\rvert_{x=\frac{L}{2}}\right)\,,
\end{align}
where we used the condition on the discontinuity of the derivative in the second line. Hence, by substituting the expressions (\ref{eq:fs}) and (\ref{eq:solBox}), we recover formula (\ref{eq:Sbox}) given in the main text.

\bibliographystyle{iopart-num}

\providecommand{\newblock}{}

\end{document}